\documentclass[12pt]{article}
\usepackage[margin=2.5cm]{geometry}  

\usepackage[margin=2.5cm]{geometry}  
\usepackage{epsfig, graphics}

\usepackage{subcaption}

\usepackage{caption}

\usepackage{amsfonts}
\usepackage{amssymb}
\usepackage{amsmath}
\usepackage[utf8]{inputenc}

\usepackage[]{natbib}
\usepackage{bookmark}

\usepackage{xcolor}
\usepackage{multirow}
\usepackage{hyperref}
\usepackage{amsmath}
\usepackage{graphicx}
\usepackage{tikz}
\usepackage{circuitikz}
\usetikzlibrary{arrows,shapes,positioning}
\usepackage{graphicx,psfrag,epsf}
\usepackage{enumerate}
\usepackage{url} 
\usepackage{siunitx}
\usepackage{float}
\usepackage{algorithm}
\usepackage{algorithmic}
\usepackage{amsmath}
\usepackage{amssymb}
\usepackage{graphicx}
\usepackage{subcaption}
\usepackage{natbib}
\usepackage{lineno}
\usepackage{dcolumn}
\usepackage{multirow}
\usepackage{comment}
\usepackage{booktabs}
\usepackage{colortbl}
\usepackage{rotating}
\usepackage{epstopdf}
\usepackage{xcolor}  
\usepackage{hyperref}
\hypersetup{
	    colorlinks=true,
    linkcolor=blue,
    urlcolor=blue,
    citecolor=blue
}
\newcommand{\nin}{\noindent}

\def\bkR{{\rm I\kern-.17em R}}

\def \1n{1\hskip -3pt \mbox{N}}

\def \Sum {\displaystyle \sum }

\newfont{\bbf}{cmbx12 scaled 1435}

\begin{document}
\setlength{\baselineskip}{.26in}
\thispagestyle{empty}
\renewcommand{\thefootnote}{\fnsymbol{footnote}}
\vspace*{0cm}
\begin{center}

\setlength{\baselineskip}{.32in}
{\bbf Generalized Covariance Estimator under Misspecification
}\\

\vspace{0.5in}

\large{ Aryan Manafi Neyazi }\footnote{University of Guelph, e-mail: {\it aneyazi@uoguelph.ca }\\The author thanks Joann Jasiak, Christian Gourieroux, Antoine Djogbenou, and the participants of the first  Non-Causal Econometrics workshop, CFE-CMStatistics 2025, Statistics Society of Canada 2026 Annual Meeting, Canadian Economics Association Annual Meeting, and 46th International Symposium on Forecasting for their helpful comments.}

\setlength{\baselineskip}{.26in}

 \vspace{0.4in}

This version: \today\\

\medskip

\vspace{0.3in}
\begin{minipage}[t]{12cm}
\small
\begin{center}
Abstract \\
\end{center}

This paper investigates the properties of the Generalized Covariance (GCov) estimator under misspecification with application to processes with local explosive patterns, such as causal-noncausal processes. We show that GCov is consistent and has an asymptotically Normal distribution under misspecification. Then, we construct GCov-based  Wald-type and score-type tests to test one specification against the other, all of which follow a $\chi^2$ distribution. We validate the finite-sample performance of the proposed estimators and tests in the context of causal-noncausal models. Finally, we provide applications of the noncausal model to the final energy demand commodity index.

\bigskip

\end{minipage}

\end{center}
\renewcommand{\thefootnote}{\arabic{footnote}}

\newpage

\section{Introduction}
Mixed causal-noncausal autoregressive (MAR) processes have become a standard tool for modeling time series with speculative bubbles, spikes, and other locally explosive patterns. By allowing the observed series to depend on both past and future values of a non-Gaussian i.i.d.\ error, MAR models generate stationary trajectories with recurrent bubble episodes [\cite{fries2019mixed}, \cite{gourieroux2017noncausal}, \cite{hecq2020mixed}], and they have been applied to forecasting during such episodes [\cite{hecq2021forecasting}, \cite{de2025forecasting}, \cite{hecq2025non}, \cite{giancaterini2025inference}], to bubble detection and dating [\cite{giancaterini2025bubble}, \cite{blasques2025novel}], and to portfolio construction [\cite{hall2024modelling}, \cite{giancaterini2025regularized}]. At the same time, it is well documented that estimation and model selection in this class are delicate. The approximate likelihood is typically multimodal, so that estimators may concentrate around spurious modes [\cite{bec2020mixed}]; identification of the noncausal component rests on non-Gaussianity, and inference is sensitive to the assumed error distribution [\cite{fries2019mixed}]; and, most importantly for this paper, \cite{gourieroux2018misspecification} show that standard order selection tools, such as information criteria and estimation under a misspecified error distribution, are unreliable for allocating a given total autoregressive order $p=r+s$ between lags $r$ and leads $s$. Since MAR($r,s$) specifications with the same total order but different lead-lag allocations are separate, non-nested models, choosing among them is intrinsically a problem of hypothesis testing under misspecification: whichever model is false, its estimator does not converge to a true parameter value but to a pseudo-true value.

This paper develops the asymptotic theory of the Generalized Covariance (GCov) estimator under parametric misspecification and builds distribution-free model selection tests on it, with MAR processes as the leading application. To position our contribution precisely, it is useful to recall what the existing GCov literature provides. \cite{gourieroux2017noncausal} introduce a semi-parametric estimator of noncausal VAR processes based on minimizing linear and nonlinear autocovariances of the residuals with a diagonal weighting of the covariance matrix. \cite{gourieroux2022generalized} propose the GCov estimator in its general form, with full covariance weighting, and establish its consistency, asymptotic normality, and semi-parametric efficiency, together with an asymptotic $\chi^2$ specification test, under correct specification of the parametric model. \cite{jasiak2023gcov} analyze this specification test under local alternatives and propose a bootstrap implementation valid for a larger class of estimators; \cite{giancaterini2025regularized} extend the estimator to high-dimensional settings through ridge regularization; and \cite{giancaterini2025bubble} employ the methodology for bubble detection. All of these results are derived under the assumption that the fitted model is correctly specified. What is missing from this literature, and what this paper supplies, is the behavior of the GCov estimator when the fitted model is wrong, which is precisely the situation faced on at least one side of any model selection exercise. The closest antecedent is \cite{gourieroux2018misspecification}, who study misspecification of the noncausal order, but they do so for pseudo-maximum likelihood estimators based on a parametric family of error distributions: their binding functions are available only for special cases, depend on the assumed distribution, and can be discontinuous in the true parameter. By working with the semi-parametric GCov estimator, we obtain a misspecification theory and closed-form binding functions that require no distributional assumption, which is essential in the MAR context where the error distribution is unknown and non-Gaussianity is itself the identifying restriction.

Our contributions are as follows. First, extending the misspecification framework of \cite{white1982maximum} and \cite{gourieroux1983testing} to the GCov setting, we show that under a misspecified parametric model the GCov estimator remains consistent for the pseudo-true value of the parameter, defined through a binding function in the spirit of \cite{sawa1978information}, and is asymptotically normal with a sandwich-form variance. We also derive the joint asymptotic distribution of the estimators obtained from the correctly specified and misspecified models. Second, based on these results we construct GCov-based Wald-type and score-type model selection tests in the tradition of \cite{gourieroux1982likelihood}, with asymptotic $\chi^2$ null distributions, covering nested, strictly non-nested, and overlapping non-nested hypotheses.\footnote{This places our tests within the broad literature on non-nested testing, including the Cox test [\cite{cox1961tests}, \cite{cox1962further}], the J and JA tests [\cite{davidson1981several}, \cite{fisher1981alternative}], encompassing tests [\cite{gourieroux1983testing}, \cite{mizon1986encompassing}], and the Vuong test [\cite{vuong1989likelihood}, \cite{shi2015nondegenerate}]; see \cite{gourieroux1995testing} and \cite{pesaran2001non} for surveys.} Because the binding function is rarely available in closed form, we provide feasible versions of the tests based on simulated pseudo-true values, adapting the indirect inference approach of \cite{gourieroux1993indirect}, so that the tests remain operational without any distributional assumption. In addition, we treat misspecification of the GCov objective function itself: the choice of the residual transformations $a(\cdot)$ and of the number of lags $H$ is a genuine source of estimation risk for the practitioner, and we show how the same testing framework can be used to compare competing sets of transformations, including regularized versions of the estimator, thereby extending the nested comparison of \cite{gourieroux2023generalized} to non-nested transformation sets. Third, we specialize the theory to causal-noncausal processes. We derive closed-form asymptotic binding functions for MAR($r,s$) models under lead-lag misspecification of any order, extending the special cases in \cite{gourieroux2018misspecification} and the pure causal representation of \cite{hecq2022spectral}, and we show that they exhibit a root-flipping structure: the unconstrained GCov estimator of a misspecified MAR model recovers the true residuals by inverting roots across the unit circle. An immediate consequence of this ``flip problem'' is that the unconstrained estimator cannot discriminate between lead-lag allocations, since the models reproduce each other's pseudo-true values and the consistency condition $b(a(\beta_0))\neq\beta_0$ of \cite{gourieroux1983testing} fails. We therefore estimate the models subject to the stationarity restrictions on the roots of the lag and lead polynomials, translated into inequality constraints on the autoregressive coefficients through the stability conditions of \cite{jury1964theory}, and we show that the constrained estimation restores $b(a(\beta_0))\neq\beta_0$ and, with it, the consistency of the proposed tests.

We should be explicit about the scope of the framework. Our null hypothesis maintains that the true data generating process belongs to one of the two competing classes, say $M_1$. Under this maintained assumption, the tests discriminate consistently between $M_1$ and a (non-)nested alternative class $M_2$. When the null is rejected, however, the tests alone cannot reveal whether the truth lies in $M_2$ or outside both classes, the scenario depicted in panel (c) of Figure \ref{figure 2}. In practice we therefore recommend, and follow in our empirical application, a two-step strategy: the maintained class is first vetted by the GCov specification test of \cite{gourieroux2022generalized}, which checks the i.i.d.\ property of its residuals without reference to $M_2$, and the model selection test is then applied to the surviving specifications. A formal treatment of the case in which both classes are misspecified, in the spirit of encompassing tests [\cite{mizon1986encompassing}, \cite{gourieroux1995testing}] or of the model comparison approach of \cite{vuong1989likelihood}, requires ranking two pseudo-true fits rather than testing a true one, and is left for future research.

We evaluate the finite-sample performance of the estimator and the tests in a dedicated Monte Carlo study built around MAR models. The experiments document the accuracy of the GCov pseudo-true values relative to their closed-form binding functions, including the absence of the discontinuities that arise for pseudo-maximum likelihood in \cite{gourieroux2018misspecification}; the empirical size of the Wald-type tests across nested, non-nested, and overlapping designs; and power curves obtained by moving systematically away from the null. Finally, we apply the methodology to the monthly Producer Price Index for final energy demand. The constrained GCov estimator selects a MAR(1,1) specification, and the comparison between constrained and unconstrained estimation illustrates, on real data, the root-flipping behavior predicted by the theory.

\textbf{Outline:} The rest of the paper is organized as follows. Section 2 reviews the GCov estimator, develops its asymptotic properties under misspecification, and constructs the Wald-type and score-type model selection tests together with their feasible simulated versions, including the treatment of misspecification in the GCov objective function. Section 3 specializes the results to causal-noncausal processes: it derives the closed-form binding functions of misspecified MAR models, documents the resulting flip problem, and introduces the constrained estimation based on the Jury conditions, which restores the consistency of the tests. Section 4 presents the Monte Carlo study of the pseudo-true values and of the size and power of the proposed tests. Section 5 contains the empirical application to the Producer Price Index for final energy demand. Section 6 concludes. Proofs and additional technical details are gathered in the Appendix.

\section{GCov Under Misspecification}

The goal of this section is to develop the asymptotic properties of the semi-parametric GCov estimator under parametric model misspecification. Based on these properties, we then construct model selection tests.

\subsection{The Generalized Covariance (GCov) Estimator}

Compared to fully parametric approaches, semi-parametric methods such as the Generalized Covariance estimator offer several advantages for estimating the coefficients of noncausal processes. \cite{gourieroux2017noncausal, gourieroux2022generalized} propose the Generalized Covariance (GCov) estimator, which is consistent and asymptotically normally distributed with a known variance under the correct specification of the parametric model and of the non-parametric part of the estimator, based on linear and nonlinear transformations of the residuals.

Let us consider a stationary process within the following semi-parametric framework:
\begin{equation}\label{eq:semi-parametric}
    g(\tilde{Y}_t; \theta) = u_t,
\end{equation}
\nin where $\tilde{Y}_t = (Y_t, Y_{t-1}, \ldots)$ and $(u_t)$ is an i.i.d.\ sequence. The function $g$ is known, while $\theta$ is an unknown parameter vector. The GCov estimator of $\theta$ is defined as
\begin{equation}\label{eq:gcov}
    \hat{\theta}_T(H) = \arg\min_{\theta} \sum_{h=1}^{H} \mathrm{Tr}[R_a^2(h, \theta)],
\end{equation}
\nin where
\begin{equation}\label{eq:R2}
    R_a^2(h, \theta) = \hat{\Gamma}^a(h; \theta)\, \hat{\Gamma}^a(0; \theta)^{-1} \hat{\Gamma}^a(h; \theta)'\, \hat{\Gamma}^a(0; \theta)^{-1}.
\end{equation}
Here, $\hat{\Gamma}^a(h; \theta)$ denotes the sample covariance matrix between $a\big(g(\tilde{Y}_t; \theta)\big)$ and $a\big(g(\tilde{Y}_{t-h}; \theta)\big)$, where $a(\cdot)$ collects $K$ linear and nonlinear transformations of the residuals.

The GCov estimator is well suited to nonlinear models in non-Gaussian frameworks. In particular, it has recently been used to estimate the parameters of causal-noncausal models [\cite{gourieroux2022generalized}, \cite{jasiak2023gcov}]. The univariate mixed causal-noncausal autoregressive process of orders $r$ and $s$, denoted MAR($r,s$), is defined by
\begin{equation} \label{eq:MAR} 
    \Phi(L)\Psi(L^{-1})y_t = \epsilon_t,
\end{equation}
where the error term $(\epsilon_t)$ is a non-Gaussian i.i.d.\ sequence. The non-Gaussianity assumption identifies the noncausal part from the causal part. The lag polynomial $\Phi(L)$ of order $r$ is backward-looking, whereas the lead polynomial $\Psi(L^{-1})$ of order $s$ is forward-looking and distinguishes the model from the traditional pure causal autoregression. The model is characterized by the roots of the causal and noncausal polynomials, which lie outside and inside the unit circle, respectively.

\medskip
\nin \textbf{Example 1:} A MAR(1,1) model fitted to $y_t$ is defined as
\[(1 - \phi L)(1 - \psi L^{-1})y_t = \epsilon_t,\]
\nin where the errors $\epsilon_t$ are i.i.d.\ with $E(|\epsilon_t|^{\delta}) < \infty$ for some $\delta > 0$, and the autoregressive coefficients $\phi$ and $\psi$ are strictly less than one in absolute value. The parameter vector is $\theta = (\phi, \psi)'$.

\subsection{Competing Specification Families}\label{subsec:families}

\begin{figure}[ht!]
    \centering
    \begin{subfigure}[b]{0.45\textwidth}
        \centering
        \includegraphics[width=\linewidth]{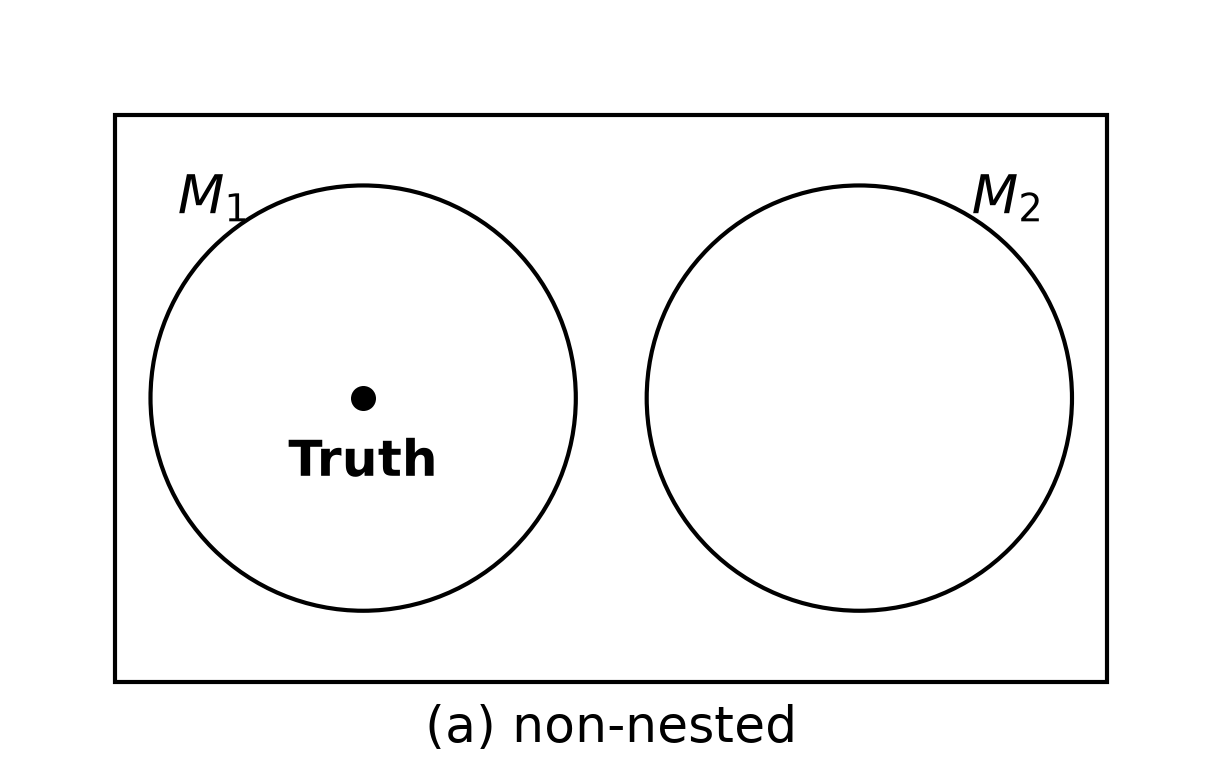}
        \caption{non-nested, truth in $M_1$}
        \label{2a}
    \end{subfigure}
    \hspace{10pt}
    \begin{subfigure}[b]{0.45\textwidth}
        \centering
        \includegraphics[width=\linewidth]{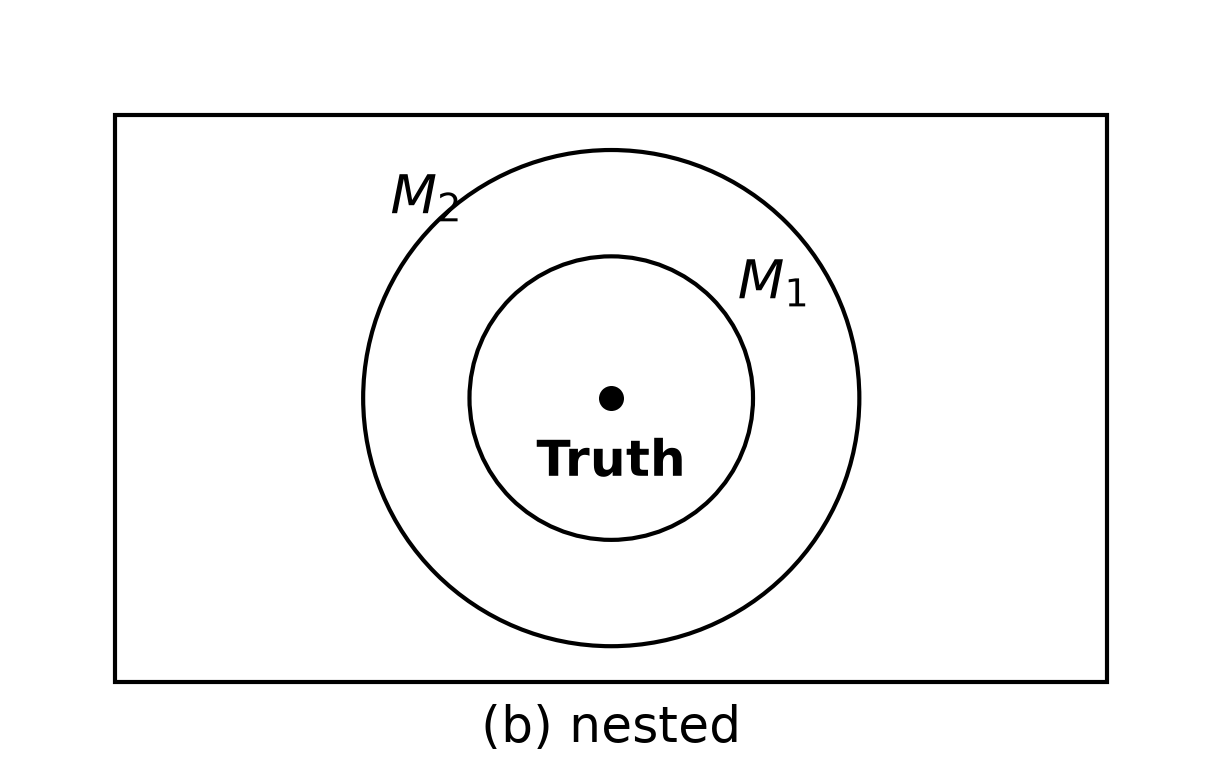}
        \caption{nested}
        \label{2b}
    \end{subfigure}

    \vspace{10pt}

    \begin{subfigure}[b]{0.45\textwidth}
        \centering
        \includegraphics[width=\linewidth]{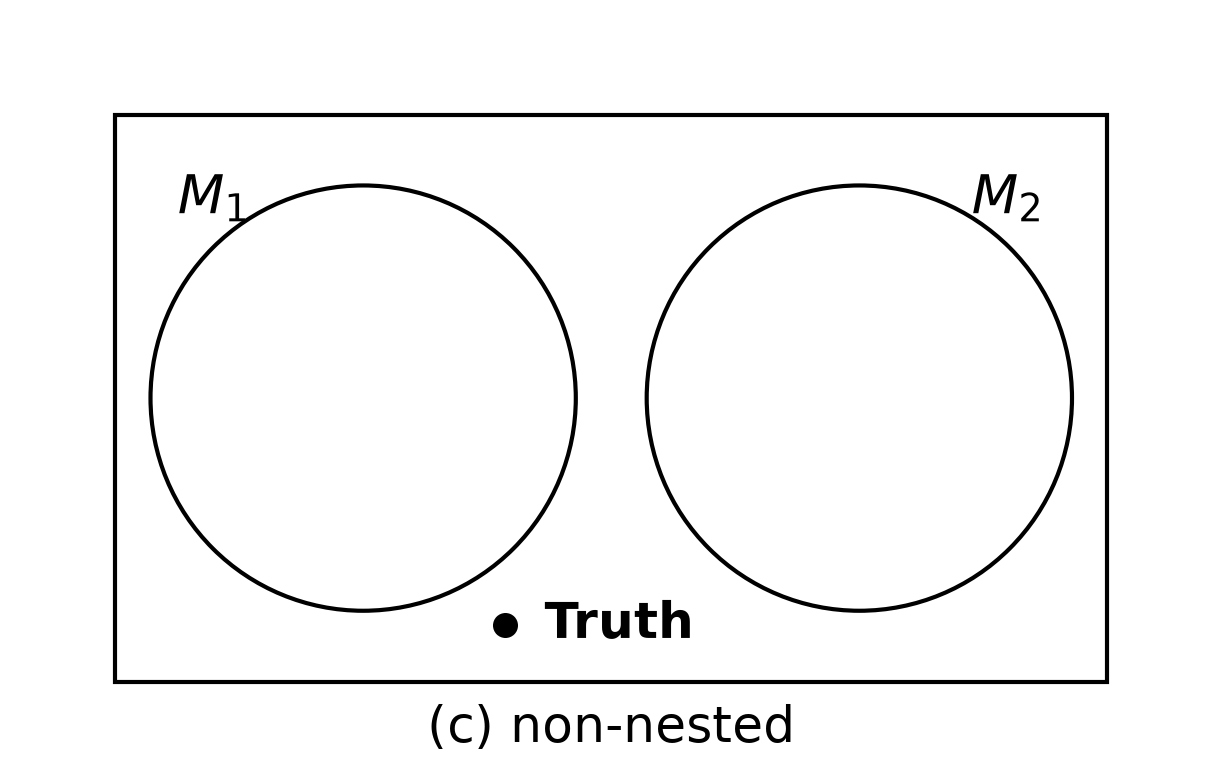}
        \caption{non-nested, truth outside both}
        \label{2c}
    \end{subfigure}
    \hspace{10pt}
    \begin{subfigure}[b]{0.45\textwidth}
        \centering
        \includegraphics[width=\linewidth]{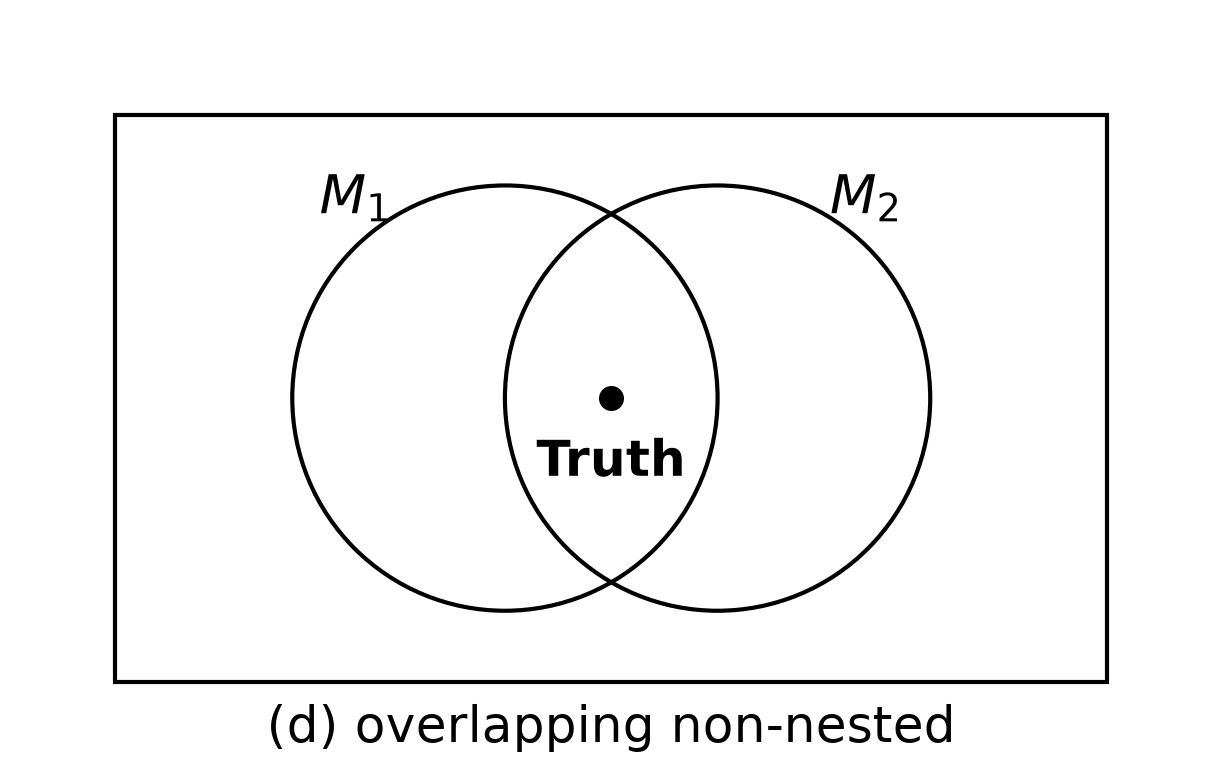}
        \caption{overlapping non-nested}
        \label{2d}
    \end{subfigure}

    \caption{Relative positions of the specification families and of the truth.}
    \label{figure 2}
\end{figure}

This study considers two specification families, denoted $M_1$ and $M_2$. These families can represent model spaces or lag lengths. Two aspects matter for the analysis: the position of the specification families relative to each other, and the position of the truth relative to them. Regarding the first aspect, three configurations can arise. First, $M_1$ and $M_2$ can be non-nested, meaning that neither family can be obtained from the other [Figure \ref{2a}]. Second, one family can be nested within the other [Figure \ref{2b}]. Third, the two families can overlap; following \cite{liao2020nondegenerate}, we call this case overlapping non-nested [Figure \ref{2d}].

\cite{gourieroux2022generalized} propose a portmanteau test based on the GCov estimation, which has an asymptotic chi-square distribution. Consider the objective function we minimize in \ref{eq:gcov} at the estimated parameter $\hat{\theta}$:
\begin{equation}
\label{L}
    L_T( \hat{\theta}_T,H) =  \sum_{h=1}^{H} Tr \bigl[ \hat{\Gamma}^a(h;  \hat{\theta}_T) \hat{\Gamma}^a(0;  \hat{\theta}_T)^{-1}\hat{\Gamma}^a(h;  \hat{\theta}_T)' \hat{\Gamma}^a(0; \hat{\theta}_T)^{-1} \bigr].
\end{equation}
Then, for the null hypothesis of 
$$H_0 : \{\Gamma^a_0 (h,\hat{\theta}_T) =0, \; h=1,...,H\},$$
we have 
\begin{equation}
    \hat{\xi}_T(H) =T L_T( \hat{\theta}_T,H),
\end{equation}
\nin which has a chi-square distribution with degrees of freedom equal to $H(Kn)^2 -dim(\theta)$, where K is the number of linear and non-linear transformations, and n is the number of variables.

\cite{jasiak2023gcov} extend the GCov test in several ways. First, they develop the asymptotic analysis of the GCov test for local alternatives and demonstrate that the test exhibits an asymptotically non-centered chi-square distribution if deviations from the null are local. Second, they propose a bootstrap GCov test that allows the use of estimators other than GCov to estimate the model's parameters. 

Regarding the second aspect, the position of the truth relative to the specification families determines whether we are under correct specification or under misspecification. While the truth is unknown, models and tests typically rely on assumptions about its position. The truth may lie outside both families, so that both are misspecified [Figure \ref{2c}]; tests have been developed to determine which family is closer to the truth in this case [\cite{vuong1989likelihood} and \cite{gourieroux1995testing}]. In the overlapping non-nested configuration, the truth may lie in the overlap, which creates an identification issue since the truth cannot be attributed to either family [Figure \ref{2d}]. Alternatively, in the nested configuration (say $M_2$ nested in $M_1$), the truth may lie inside $M_2$, in which case $M_1$ is overfitting [Figure \ref{2b}]. In what follows, we assume that $M_1$ and $M_2$ are strictly non-nested and that the truth lies in one of them, without loss of generality in $M_1$ [Figure \ref{2a}]. This assumption allows us to derive the asymptotic distributions of the tests under a true null hypothesis. We relax this assumption later and allow for the nested case.

\subsection{Misspecification in the Parametric Model} 

Consider the following non-nested model spaces:
$$M_1 : g(\tilde{Y}_t; \theta) = u_t,\;\; M_2 : h(\tilde{Y}_t; \beta) = v_t,$$
\nin where $g$ and $h$ are known functions satisfying the assumptions of the previous subsection, the two families are strictly non-nested, and $\tilde{Y}_t= (Y_t, Y_{t-1},\ldots, Y_{t-p})$. Without loss of generality, assume that the true model space is $M_1$ and that the true distribution of the errors is $f_0$. The pseudo-true distribution of the errors under the misspecified family $M_2$ is denoted $l_0(\theta_0,f_0)$. The parameters of interest are $\theta \in \Theta$ and $\beta \in \mathcal{B}$.

\medskip
\nin \textbf{Assumption 1:} 

\nin (i) The process $(Y_t)$ is strictly stationary and the errors are i.i.d.\ with true distribution $f_0$.

\nin (ii) The functions $g$ and $h$ are invertible with respect to $Y_t$ and differentiable for all $\theta \in \Theta$ and $\beta \in \mathcal{B}$, respectively.

\medskip
\nin \textbf{Assumption 2:} The pseudo-true value of the parameter, $b(\theta_0,f_0)$, and the finite-sample pseudo-true value of the parameter, $b_T(\theta_0,f_0)$, exist, are unique, and belong to the compact set $\mathcal{B}$.

\medskip
\nin \textbf{Assumption 3:} The binding function $b(\cdot)$ is one-to-one and the matrix $\frac{\partial b}{\partial \theta'}(\theta_0,f_0)$ has full column rank.

\medskip
\nin \textbf{Assumption 4:} The matrix 
$$\sum_{h=1}^{H} \frac{\partial \mathrm{Tr}[R_a^2(h, \beta)] }{\partial \beta } [b(\theta_0,f_0)]\, \frac{\partial \mathrm{Tr}[R_a^2(h, \beta)] }{\partial \beta' } [b(\theta_0,f_0)]$$ 
\nin is positive semi-definite, and the matrix
$$\sum_{h=1}^{H} \frac{\partial^2 \mathrm{Tr}[R_a^2(h, \beta)] }{\partial \beta \partial \beta'}[b(\theta_0,f_0)]$$ 
\nin is positive definite.

\medskip
Assumption 2 guarantees the existence and uniqueness of the pseudo-true value of the parameter. Assumption 3 follows from the differentiability of the binding function. Assumption 4 ensures that the asymptotic variance is well behaved.

Since the true model space is $M_1$, the function $g(\cdot)$ is the true model, the true value of the parameter is $\theta_0$, and the estimator $\hat{\theta}_T$ converges to $\theta_0$ asymptotically [\cite{gourieroux2022generalized}]. In contrast, when the model in $M_2$ is fitted, the estimator $\hat{\beta}_T$ converges asymptotically to the pseudo-true value $b(\theta_0,f_0)$. The estimators are obtained by minimizing the GCov objective function under each model:
\begin{equation}
    \hat{\theta}_T(H) = \arg\min_{\theta} \sum_{h=1}^{H} \mathrm{Tr}[R_a^2(h, \theta)],
\end{equation}
and
\begin{equation}
    \hat{\beta}_T(H) = \arg\min_{\beta} \sum_{h=1}^{H} \mathrm{Tr}[R_a^2(h, \beta)].
\end{equation}

\medskip
\nin \textbf{Proposition 1:} Under Assumptions 1 to 4 and the regularity conditions in Appendix A.1, the GCov estimator is consistent for the pseudo-true value of the parameter and satisfies
\begin{equation}\label{eq:prop1dist}
    \sqrt{T}\big(\hat{\beta}_T- b(\theta_0,f_0)\big) \xrightarrow{d} N\big(0,\Omega^a_{22}(H,b(\theta_0,f_0))\big),
\end{equation}
\nin where
$$\Omega^a_{22}(H,b(\theta_0,f_0))=J^a_{22}(H,b(\theta_0,f_0))^{-1}I^a_{22}(H,b(\theta_0,f_0))J^a_{22}(H,b(\theta_0,f_0))^{-1},$$
$$ J^{a}_{22} (H, b(\theta_0,f_0)) = \sum_{h=1}^{H} \frac{\partial^2 \mathrm{Tr}[R_a^2(h, \beta)] }{\partial \beta \partial \beta'}[b(\theta_0,f_0)],$$
$$ I^{a}_{22} (H,b(\theta_0,f_0))= \sum_{h=1}^{H} \frac{\partial \mathrm{Tr}[R_a^2(h, \beta)] }{\partial \beta } [b(\theta_0,f_0)]\, \frac{\partial \mathrm{Tr}[R_a^2(h, \beta)] }{\partial \beta' } [b(\theta_0,f_0)].$$
\nin \textbf{Proof:} See Appendix A.

\medskip
\nin \textbf{Remark 1:} Under correct specification, Proposition 1 reduces to the asymptotic properties of the GCov estimator established in \cite{gourieroux2022generalized}.

\medskip
Comparing the asymptotic distributions of the GCov estimator under correct specification and under misspecification, both are asymptotically normal; however, the semi-parametric efficiency property is lost under misspecification. Based on Proposition 1 and following the approach of \cite{gourieroux1983testing} for the PML estimator, we obtain the joint distribution in Corollary 1.

\medskip
\nin \textbf{Corollary 1:} If model $M_1$ is correctly specified and model $M_2$ is misspecified, then by Proposition 1 and the asymptotic normality of the GCov estimator under correct specification, the vector 
\[
\sqrt{T}
\begin{pmatrix}
    \hat{\theta}_T - \theta_0 \\
    \hat{\beta}_T- b(\theta_0,f_0)
\end{pmatrix}
\]
\nin is asymptotically normal with mean zero and variance 
$$\Omega^a(H,\theta_0,b(\theta_0,f_0))=J^a (H,\theta_0,b(\theta_0,f_0))^{-1} I^a(H,\theta_0,b(\theta_0,f_0))J^a (H,\theta_0,b(\theta_0,f_0))^{-1},$$
\nin where 
\[
J^a (H,\theta_0,b(\theta_0,f_0))=
\begin{pmatrix}
    J^{a}_{11} (H,\theta_0) & 0 \\
    0 & J^{a}_{22} (H,b(\theta_0,f_0))
\end{pmatrix},
\]
\[
I^a (H,\theta_0,b(\theta_0,f_0))=
\begin{pmatrix}
    I^{a}_{11} (H,\theta_0) & I^{a}_{12} (H,\theta_0,b(\theta_0,f_0)) \\
    I^{a}_{21} (H,\theta_0,b(\theta_0,f_0)) & I^{a}_{22} (H,b(\theta_0,f_0))
\end{pmatrix},
\]
$$ J^{a}_{11} (H, \theta_0) = \sum_{h=1}^{H} \frac{\partial^2 \mathrm{Tr}[R_a^2(h, \theta)] }{\partial \theta \partial \theta'}[\theta_0],$$
$$ I^{a}_{11} (H,\theta_0)= \sum_{h=1}^{H} \frac{\partial \mathrm{Tr}[R_a^2(h, \theta)] }{\partial \theta } [\theta_0]\, \frac{\partial \mathrm{Tr}[R_a^2(h, \theta)] }{\partial \theta '} [\theta_0],$$
$$ I^{a}_{12} (H,\theta_0,b(\theta_0,f_0))= \sum_{h=1}^{H} \frac{\partial \mathrm{Tr}[R_a^2(h, \theta)] }{\partial \theta } [\theta_0]\, \frac{\partial \mathrm{Tr}[R_a^2(h, \beta)] }{\partial \beta' } [b(\theta_0,f_0)] = I^{a}_{21}  (H,\theta_0,b(\theta_0,f_0))'. $$

\medskip
Proposition 1 and Corollary 1 characterize the asymptotic distribution of $\hat{\beta}_T$ around the pseudo-true value $b(\theta_0,f_0)$, which is usually unknown to the researcher. It is therefore useful to distinguish the four binding-function objects used in the sequel: the pseudo-true value $b(\theta_0,f_0)$, evaluated at the true parameter and the true error distribution; its finite-sample counterpart $b_T(\theta_0,f_0)$; the asymptotic pseudo-true value $b(\hat{\theta},\hat{f})$, in which the true quantities are replaced by their estimates; and the finite-sample pseudo-true value $b_T(\hat{\theta},\hat{f})$. When $b(\cdot)$ is written with a single argument, the error distribution is held fixed implicitly.

\medskip
\nin \textbf{Proposition 2:} Under Assumptions 1 to 4, the GCov estimator is asymptotically normal around the asymptotic pseudo-true value $b(\hat{\theta},\hat{f})$ and around the finite-sample pseudo-true value $b_T(\hat{\theta},\hat{f})$, with respective variances
\begin{equation*}
        \Omega^a_{A} = {J^a_{22}}^{-1} [ I^a_{22}  - I^a_{21}  {I^a_{11}} ^{-1} I^a_{12}] {J^a_{22}}^{-1} ,
\end{equation*}
\nin and 
\begin{equation*}
        \Omega^a_{F} = {J^a_{22}}^{-1} [ I^{a*}_{22}  - I^a_{21}  {I^a_{11}} ^{-1} I^a_{12}] {J^a_{22}}^{-1} ,
\end{equation*}
where
\[
 I^{a*}_{22}= E_{\theta_0}\!\left[ \sum_{h=1}^{H}\left( \frac{\partial \mathrm{Tr}[R_a^2(h, \beta)] }{\partial \beta } [b(\theta_0,f_0)] - E_{\theta_0} \frac{\partial \mathrm{Tr}[R_a^2(h, \beta)] }{\partial \beta } [b(\theta_0,f_0)] \right) \right.
 \]
 \[
 \left. \times \sum_{h=1}^{H}\left(\frac{\partial \mathrm{Tr}[R_a^2(h, \beta)] }{\partial \beta } [b(\theta_0,f_0)] - E_{\theta_0} \frac{\partial \mathrm{Tr}[R_a^2(h, \beta)] }{\partial \beta } [b(\theta_0,f_0)]  \right)' \right].
\]
\nin \textbf{Proof:} See Appendix A.

\medskip
Even the closed form of the finite-sample pseudo-true value $b_T(\hat{\theta},\hat{f})$ may not be available. We therefore present a simulation approach, proposed by \cite{gourieroux1993indirect}, that provides an asymptotically consistent estimator of the pseudo-true value. The approach consists of the following steps:

\nin - Estimate the parameter under the correct specification, yielding the estimate $\hat{\theta}$ and the fitted residuals $\hat{u}_t$.

\nin - Resample the residuals to obtain $\hat{u}^s_t$ for $s=1,2,\ldots, S$.

\nin - Generate $y^s_t=g^{-1}(\hat{\theta},\hat{u}^s_t)$.

\nin - For each simulated series $y^s_t$, fit the misspecified model $h(\cdot)$ and obtain the estimate $\hat{\beta}^s$.

\nin - Compute
\begin{equation}
    b_{T,S}(\hat{\theta},\hat{f})= \frac{1}{S} \sum_{s=1}^S \hat{\beta}^s.
    \label{b_T,S}
\end{equation}
\nin This simulation path is computationally demanding. \cite{gourieroux1993indirect} suggest an alternative: instead of estimating $\hat{\beta}^s$ for $s=1,\dots,S$, one can simulate a single time series of length $TS$ and estimate $b_{TS}(\hat{\theta},\hat{f})$, which is asymptotically equivalent to $b_{T,S}(\hat{\theta},\hat{f})$.

\medskip
\nin \textbf{Remark 2:} Based on the simulated finite-sample pseudo-true value $b_{T, S}(\hat{\theta},\hat{f})$ and in the pure time series framework [\cite{gourieroux1995testing}], the quantities $\sqrt{T}\big(\hat{\beta}_T - b_{T,S}(\hat{\theta},\hat{f}) \big)$ and $\sqrt{T}\big(\hat{\beta}_T - b_{TS}(\hat{\theta},\hat{f}) \big)$ are asymptotically equivalent and asymptotically normal with mean zero and variance-covariance matrix 
$$\Omega^a_{S}=\left( 1 + \frac{1}{S} \right) {J^a_{22}}^{-1}  I^{a*}_{22}\, {J^a_{22}}^{-1}.$$

\subsection{Model Selection Based on GCov}\label{subsec:tests}

A model can be considered correctly specified if the null hypothesis of i.i.d.\ residuals is not rejected by the GCov specification test, and misspecified if that null is rejected. This argument is sensitive to the number of lags included in the GCov objective function and to the transformations considered. Let $M_1$ be the correct specification and $M_2$ the misspecified model space. Across different transformations, numbers of transformations, and lags, we expect $M_1$ to always appear correctly specified; however, $M_2$ may be falsely accepted as correctly specified when the transformations are non-informative. 

The GCov-based specification test was proposed by \cite{gourieroux2022generalized}, and \cite{jasiak2023gcov} investigated its properties under local alternatives. Here, we extend this line of work to a broader set of model selection tools, allowing one specification to be tested against another when the models are nested, overlapping, or non-nested.

We first consider the Wald-type approach introduced by \cite{gourieroux1983testing} and \cite{white1982maximum}. This approach relies on the concept of the pseudo-true parameter value, originally developed by \cite{sawa1978information} and \cite{white1982maximum}, which subsequently played a central role in the development of encompassing tests [\cite{mizon1986encompassing}, \cite{gourieroux1995testing}]. The regularity conditions in this subsection follow \cite{gallant1980statistical} and \cite{burguete1980unification}.

We introduce a Wald-type test based on the GCov estimator for testing between two separate model spaces, in the framework of Subsection \ref{subsec:families}. The GCov estimator is attractive in this context because of its advantages under non-Gaussian i.i.d.\ errors. The test is particularly useful for selecting the noncausality order of mixed autoregressive models.

\medskip
\nin \textbf{Corollary 2:} Under Assumptions 1 to 4, and based on Propositions 1 and 2, we propose the following GCov-based Wald-type test statistics:
\begin{equation}
    \xi_{T}^{W1}=T\big(\hat{\beta}-b(\hat{\theta},\hat{f})\big)'\hat{\Omega}^{a^{-1}}_{A}\big(\hat{\beta}-b(\hat{\theta},\hat{f})\big),
    \label{GET1}
\end{equation}
\begin{equation}
    \xi_{T}^{W2}=T\big(\hat{\beta}-b_T(\hat{\theta},\hat{f})\big)'\hat{\Omega}^{a^{-1}}_{F}\big(\hat{\beta}-b_T(\hat{\theta},\hat{f})\big),
    \label{GET2}
\end{equation}
\begin{equation}
    \xi_{T}^{W3}=T\big(\hat{\beta}-b_{T,S}(\hat{\theta},\hat{f})\big)'\hat{\Omega}^{a^{-1}}_{S}\big(\hat{\beta}-b_{T,S}(\hat{\theta},\hat{f})\big),
    \label{GET3}
\end{equation}
\nin all of which are asymptotically $\chi^2$-distributed with degrees of freedom $d_{1}$, $d_{2}$, and $d_{3}$, equal to the ranks of the population matrices $\Omega^a_{A}$, $\Omega^a_{F}$, and $\Omega^a_{S}$, respectively. The proposed tests are consistent if and only if $b(a(\beta_0)) \neq \beta_0$ [\cite{gourieroux1983testing}, \cite{gourieroux1995testing}].

\nin The asymptotic distributions of $\xi_{T}^{W1}$, $\xi_{T}^{W2}$, and $\xi_{T}^{W3}$ are a direct consequence of Proposition 2 and Corollary 1.

\medskip
\nin \textbf{Remark 3:} In the nested case, these statistics reduce to the traditional GCov-based Wald test.

\medskip
Next, we propose a GCov-based score-type test. Following the approach of \cite{gourieroux1983testing} for the MLE, we define
$$\hat{\lambda}_T^{(1)}= \sum_{h=1}^{H} \frac{\partial \mathrm{Tr}[R_a^2(h, \beta)] }{\partial \beta } [b(\hat{\theta},\hat{f})],$$
\nin and
$$\hat{\lambda}_T^{(2)}= \sum_{h=1}^{H} \frac{\partial \mathrm{Tr}[R_a^2(h, \beta)] }{\partial \beta } [b_T(\hat{\theta},\hat{f})],$$
\nin and construct tests of their departure from zero. The following proposition gives the asymptotic distributions of these score functions.

\medskip
\nin \textbf{Proposition 3:} If the correct specification belongs to $M_1$, then
$$\sqrt{T}\hat{\lambda}_T^{(1)} \xrightarrow{d} N\big(0,\,I^{a}_{22}- I^a_{21} {I^a_{11}}^{-1} I^a_{12}\big),$$
\nin and
$$\sqrt{T}\hat{\lambda}_T^{(2)} \xrightarrow{d} N\big(0,\,I^{a*}_{22}- I^a_{21} {I^a_{11}}^{-1} I^a_{12}\big).$$
\nin \textbf{Proof:} See Appendix A.

\medskip
Based on these asymptotic distributions, we construct an extension of the traditional score test, which is asymptotically equivalent to the Wald-type extensions above.

\medskip
\nin \textbf{Corollary 3:} Based on Proposition 3, the statistics
\begin{equation}
    \xi_{T}^{S1}= T\, \hat{\lambda}_T^{(1)'} \left[I^{a}_{22}- I^a_{21} {I^a_{11}}^{-1} I^a_{12} \right]^{-1}  \hat{\lambda}_T^{(1)},
    \label{GST1}
\end{equation}
and
\begin{equation}
    \xi_{T}^{S2}= T\, \hat{\lambda}_T^{(2)'} \left[I^{a*}_{22}- I^a_{21} {I^a_{11}}^{-1} I^a_{12} \right]^{-1}  \hat{\lambda}_T^{(2)},
    \label{GST2}
\end{equation}
\nin are both asymptotically $\chi^2$-distributed with degrees of freedom equal to the rank of the variance matrix of the corresponding score function.

It is worth noting that all asymptotic distributions of the tests presented in this subsection are derived under the assumption that the null hypothesis is correctly specified, as is standard in the development of asymptotic theory for hypothesis testing. From a practical perspective, we therefore suggest conducting the test in both directions when comparing two competing models, in order to obtain a more informative assessment of which specification provides a better fit to the observed time series.

Nevertheless, it is always possible that neither of the parametric models under consideration is correctly specified. In such cases, the proposed framework would require a more advanced extension toward an encompassing-testing approach, along the lines of \cite{mizon1986encompassing} and \cite{gourieroux1995testing}. Such an extension would develop an encompassing test within the GCov framework rather than relying on maximum likelihood estimation. We leave this extension for future research.

\subsection{Misspecification in the GCov Objective Function} 

Consider now the case in which the parametric model is correctly specified ($M_1$), but the objective function of the GCov estimator is misspecified. One example is the objective function proposed by \cite{gourieroux2017noncausal}, which uses $\mathrm{diag}\,\hat{\Gamma}(0; \theta)$ instead of $\hat{\Gamma}(0; \theta)$. Another example is any regularization of $\hat{\Gamma}(0; \theta)$ designed to make the GCov estimator applicable in high-dimensional settings, where invertibility becomes an issue, as proposed by \cite{giancaterini2025regularized} for the case of ridge regularization. A third source is the set of transformations included in the GCov estimator: with non-informative transformations, the GCov estimator remains consistent and asymptotically normal, but its variance may be affected.

We now propose a way to choose between sets of transformations to be included in the GCov estimator. We introduce two specifications, $M_1'$ and $M_2'$, in which the parametric model is correctly specified (model \eqref{eq:semi-parametric}, i.e., $M_1$) and which differ only in the transformations used. For now, we impose no assumption on whether the transformation sets are nested or non-nested. Suppose $M_1'$ contains all the informative transformations, while $M_2'$ contains a less informative set. Proposition 3 then simplifies as in Remark 4.

\medskip
\nin \textbf{Remark 4:} The asymptotic distributions of the GCov estimator under $M_1'$ and $M_2'$ are
\begin{equation}\label{eq:rem4a}
    \sqrt{T}\big(\hat{\theta}^{M_1'}_T- \theta_0\big) \xrightarrow{d} N\big(0,\,J^{a'}_{11}(H,\theta_0)^{-1}\big),
\end{equation}
\nin and
\begin{equation}\label{eq:rem4b}
    \sqrt{T}\big(\hat{\theta}^{M_2'}_T- \theta_0\big) \xrightarrow{d} N\big(0,\,J^{a'}_{22}(H,\theta_0)^{-1}\big).
\end{equation}
\nin Consequently, all the previous results remain valid in this framework after substituting $\theta$ for $\beta$, $\theta_0$ for $b(\theta_0)$, $J^{a'}$ for $J^{a}$, and $I^{a'}$ for $I^{a}$, where $\mathrm{Tr}[R_a^2]$ is indexed by $M_1'$ and $M_2'$. Since the parametric model is correctly specified, the matrices $J^{a'}$ and $I^{a'}$ have a simpler structure, as in \cite{gourieroux2022generalized}.

\medskip
\nin \textbf{Remark 5:} Remark 4 continues to hold if $M_2'$ is identified with the RGCov estimator of the parametric model in the high-dimensional framework.

\medskip
Based on Remarks 4 and 5, we can construct Wald-type or score-type tests to choose between two sets of transformations, or between values of the regularization parameter of the RGCov estimator, along the lines of Subsection \ref{subsec:tests}. This extends Proposition 4 of \cite{gourieroux2023generalized}, which considers only the nested transformation scenario.

\section{Causal-Noncausal Autoregressive Models with Constraints}

The concept of misspecification was introduced into causal-noncausal models by \cite{gourieroux2018misspecification}, who consider a misspecified allocation of lags and leads, estimated by PML under a misspecified error distribution. An interesting feature of MAR models is that closed-form binding functions are attainable under misspecification, as shown for special cases in \cite{gourieroux2018misspecification}. In this section, we extend their work to a broader setting and relax the assumption of a known distribution by deriving the binding functions of the semi-parametric GCov estimator. Note that the binding functions of \cite{gourieroux2018misspecification} cannot be used for our model selection tests, as they are based on the PML estimator. Since MAR($r,s$) specifications with a fixed total number of lags and leads are non-nested for $r \geq 1$ and $s \geq 1$, and the existing selection criteria for MAR models are biased [\cite{gourieroux2018misspecification}], the GCov-based model selection tests provide a clear contribution.

\medskip
\nin \textbf{Remark 6:} Consider the DGP of MAR($r,s$) and the misspecified model MAR($r-q$, $s+q$), where $q$ is the order of misspecification. The roots of the lag polynomial of the correct specification are $\lambda_1^{-1},\ldots,\lambda_r^{-1}$, with $|\lambda_i|<1$, and the roots of the lead polynomial are $\gamma_1,\ldots,\gamma_s$, with $|\gamma_i|<1$. Suppose the $q$ roots of the lag polynomial that flip to the lead side are the last $q$ roots. The roots under the misspecified model are then $\lambda_1,\ldots,\lambda_{r-q}$ and $\gamma_1,\ldots,\gamma_{s+q}$, where $\gamma_{s+i}=\lambda_{r-i}$ for $i=1,\ldots,q$. The asymptotic binding functions of the pseudo-true parameters are, for $i=1,\ldots,r-q$,
$$b_{\phi_{i}}(\phi_{0,1},\ldots,\phi_{0,r},\psi_{0,1},\ldots,\psi_{0,s})= (-1)^{i+1} \sum_{j_1 =1}^{r-q} \sum_{j_1<j_2}^{r-q}\cdots\sum_{j_{i-1}<j_{i}}^{r-q} \lambda_{j_1} \lambda_{j_2}\cdots \lambda_{j_{i}},$$
\nin and, for $i=1,\ldots,s+q$,
$$b_{\psi_{i}}(\phi_{0,1},\ldots,\phi_{0,r},\psi_{0,1},\ldots,\psi_{0,s})= (-1)^{i+1} \sum_{j_1 =1}^{s+q} \sum_{j_1<j_2}^{s+q}\cdots\sum_{j_{i-1}<j_{i}}^{s+q} \gamma_{j_1} \gamma_{j_2}\cdots \gamma_{j_{i}}.$$
\nin Since any $q$ of the $r$ lag roots may flip, there are $\binom{r}{q}=\frac{r!}{(r-q)!\, q!}$ possible sets of pseudo-true parameter values under the unconstrained GCov estimator.

\medskip
Remark 6 extends the pure causal representation of MAR($r,s$) proposed by \cite{hecq2022spectral}, since the pure causal representation of a MAR($r,s$) process can be viewed as a misspecified model. Moreover, it provides closed-form binding functions for any order of misspecification, extending \cite{gourieroux2018misspecification}.

\medskip
\nin \textbf{Remark 7:} By Remark 6, when the order of misspecification $q$ is non-zero, the pseudo-true values of the parameters under the misspecified model do not lie in the region that satisfies the model's assumptions. In that case, a Wald-type or score-type test cannot be used to choose between causal and noncausal models. With a constrained GCov estimator, however, $b(a(\beta_0))\neq \beta_0$ holds, which allows the use of the proposed tests.

\medskip
\nin \textbf{Remark 8:} Consider model $M_1$: MAR($r,s$) with true parameter vector $\alpha_0$ and model $M_2$: MAR($r-q$, $s+q$) with true parameter vector $\beta_0$. If we are under the $M_1$ specification, the binding function is $b(\alpha_0)$; if we are under the $M_2$ specification, the binding function is $a(\beta_0)$. For any non-zero order of misspecification $q$, the unconstrained GCov estimator yields $b(a(\beta_0))=\beta_0$ and $a(b(\alpha_0))=\alpha_0$.

\medskip
\nin \textbf{Remark 9:} Consider $M_1$: MAR($r,s$) with $r+s=p$ and $M_2$: MAR($r',s'$) with $r'+s'=p'$ and $p<p'$. If we are under the $M_1$ specification, then $b(a(\beta_0))\neq \beta_0$.

\medskip
We now turn to the GCov estimation of constrained causal-noncausal models. To estimate the parameters $\theta$ of the MAR($r,s$) model in \eqref{eq:MAR} subject to the root conditions $|\lambda_i|<1$ and $|\gamma_i|<1$, we consider
\begin{eqnarray}
    \hat{\theta}^{\,C}_T(H)& = &\arg\min_{\theta} \sum_{h=1}^{H} \mathrm{Tr}[R_a^2(h, \theta)], \label{eq:cobj}\\
    &s.t.&  |\lambda_i|<1,\; i=1,\ldots,r, \quad |\gamma_i|<1,\; i=1,\ldots,s, \label{eq:constraint}
\end{eqnarray}
\nin where $R_a^2(h, \theta)$ is defined in \eqref{eq:R2}.

The constraint \eqref{eq:constraint} is expressed on the roots of the polynomials; however, it can be transformed into an equivalent set of constraints on $\theta$. We use the algorithm proposed by \cite{jury1964theory} to convert the root conditions into conditions on the parameters. Consider the lag polynomial of order $r$:
$$\Phi(L)=1-\phi_1L-\phi_2L^2-\cdots-\phi_r L^r,$$
\nin whose roots must lie outside the unit circle. This is equivalent to requiring that the roots of
$$L^r \Phi(L^{-1})=-\phi_r -\phi_{r-1}L-\cdots-\phi_1L^{r-1}+L^r$$
\nin lie inside the unit circle. For notational simplicity, we rewrite this polynomial as
$$F(z)=a_0+a_1z+\cdots+a_rz^r,$$
\nin where $z=L$, $a_r=1$, and $a_i=-\phi_{r-i}$ for $i=0,1,\ldots,r-1$. We then construct the matrices $X_{k}$ and $Y_{k}$ as
\begin{equation}\label{matrix:X}
    X_{k}=
    \begin{bmatrix}
        a_0   & a_1   & a_2   & \dots  & a_{k-1} \\
        0     & a_0   & a_1   & \dots  & a_{k-2} \\
        0     & 0     & a_0   & \dots  & a_{k-3} \\
        \vdots     & \vdots     & \vdots    & \vdots    & \vdots        \\
        0     & 0     & 0     & \dots  &  a_0    \\
    \end{bmatrix},
    \quad
    Y_{k}=
    \begin{bmatrix}
        a_{r-k+1}     & \dots   & a_{r-2}   & a_{r-1}  & a_{r} \\
        a_{r-k+2}     & \dots   & a_{r-1}   & a_r  & 0         \\
        a_{r-k+3}    & \dots   & a_r       & 0    & 0         \\
        \vdots     & \vdots     & \vdots    & \vdots    & \vdots        \\
        a_r     & \dots   & 0         & 0    &  0        \\
    \end{bmatrix}.
\end{equation}
\nin We write the determinants as $|X_k+Y_k|=A_k+B_k$ and $|X_k-Y_k|=A_k-B_k$, where $A_k$ and $B_k$ are the stability constants of \cite{jury1964theory}. The condition that the roots of $F(z)$ lie inside the unit circle, equivalently that the roots of $\Phi(L)$ lie outside the unit circle, is, for $r$ odd,
$$F(1)>0, \qquad F(-1)<0,$$
$$(-1)^{k(k+1)/2}(A_k \pm B_k)>0, \qquad k=2,4,6,\dots,r-1,$$
\nin and, for $r$ even,
$$F(1)>0, \qquad F(-1)>0,$$
$$(-1)^{k(k+1)/2}(A_k -B_k)>0,\qquad (-1)^{k(k+1)/2}(A_k +B_k)<0, \qquad k=1,3,5,\dots,r-1.$$
\nin The same approach applies to the lead polynomial.

\medskip
\nin \textbf{Example 2:} Consider the MAR(3,3) model
$$(1-\phi_1L-\phi_2L^2-\phi_3 L^3)(1-\psi_1L^{-1}-\psi_2 L^{-2} -\psi_3 L^{-3})y_t=\epsilon_t,$$
\nin where the roots of the lag polynomial lie outside, and those of the lead polynomial inside, the unit circle. The equivalent constraints on the parameters are:
\begin{eqnarray*}
       & -\phi_3-\phi_2-\phi_1+1>0 ,&  -\phi_3+\phi_2-\phi_1-1<0, \\
    & |-\phi_3|<1,& \phi_3^2-1<\phi_3\phi_1+\phi_2,  \\
     & -\psi_3-\psi_2-\psi_1+1>0 ,&  -\psi_3+\psi_2-\psi_1-1<0 , \\
    & |-\psi_3|<1 ,& \psi_3^2-1<\psi_3\psi_1+\psi_2. \\
\end{eqnarray*}

\color{black}
\section{Simulation Studies}

This section investigates the use of the proposed estimator and tests in causal-noncausal processes. These models satisfy the assumptions of the GCov estimator both under correct specification and misspecification. 

\subsection{Estimator and Pseudo-True Values }

In the first step, we generate MAR(0,2) and fit MAR(1,1) with the GCov estimator without any constraints on the roots to see how well it estimates the pseudo-true value of the parameter. Then we report the RMSE of the estimator of the pseudo-true estimator.

To compare the pseudo-true values of the parameters based on GCov and ML estimators, we conduct the same simulation as provided in \cite{gourieroux2018misspecification} Figure 2 by generating a noncausal AR(1) with a Cauchy error distribution and with different autoregressive coefficients from 0.1 to 0.9. The number of observations is T=1000. Then we fit a causal AR(1) as a misspecified model and report the mean of the estimator for 1000 replications in Figure 2. Comparing Figure 2 with Figure 2 of \cite{gourieroux2018misspecification} shows the advantage of using the GCov estimator in terms of not having a discontinuity in the pseudo-true value of the parameter. 
\begin{figure}[ht]
    \centering
    \includegraphics[width=0.7\linewidth]{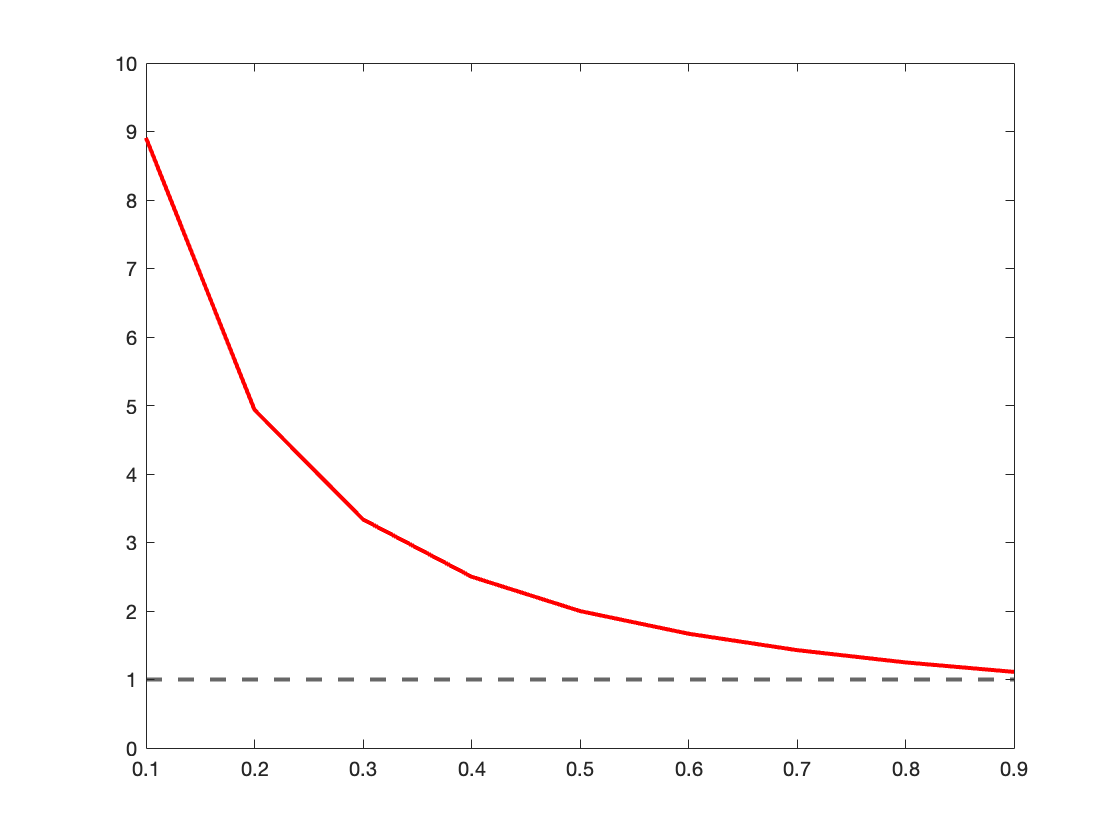}
    \caption{Mean pseudo-true value of Misspecified causal AR(1) when the correct model is noncausal AR(1) with Cauchy error distribution.}
    \label{MAR(1,1)C}
\end{figure}

To check the normality of the GCov estimator around pseudo-true values, we generate $MAR(0,1)$ with $\psi=0.3$, errors with $t(6)$ distribution, $T=2000$ observations and 1000 replications. Then we fit $MAR(0,2)$ as misspecified model. Figure 3 provides the histogrum of the estimates and their QQ-plots.

 \begin{figure}[ht!]
    \centering
    
    \begin{subfigure}[b]{0.49\textwidth}
        \centering
        \includegraphics[width=\textwidth]{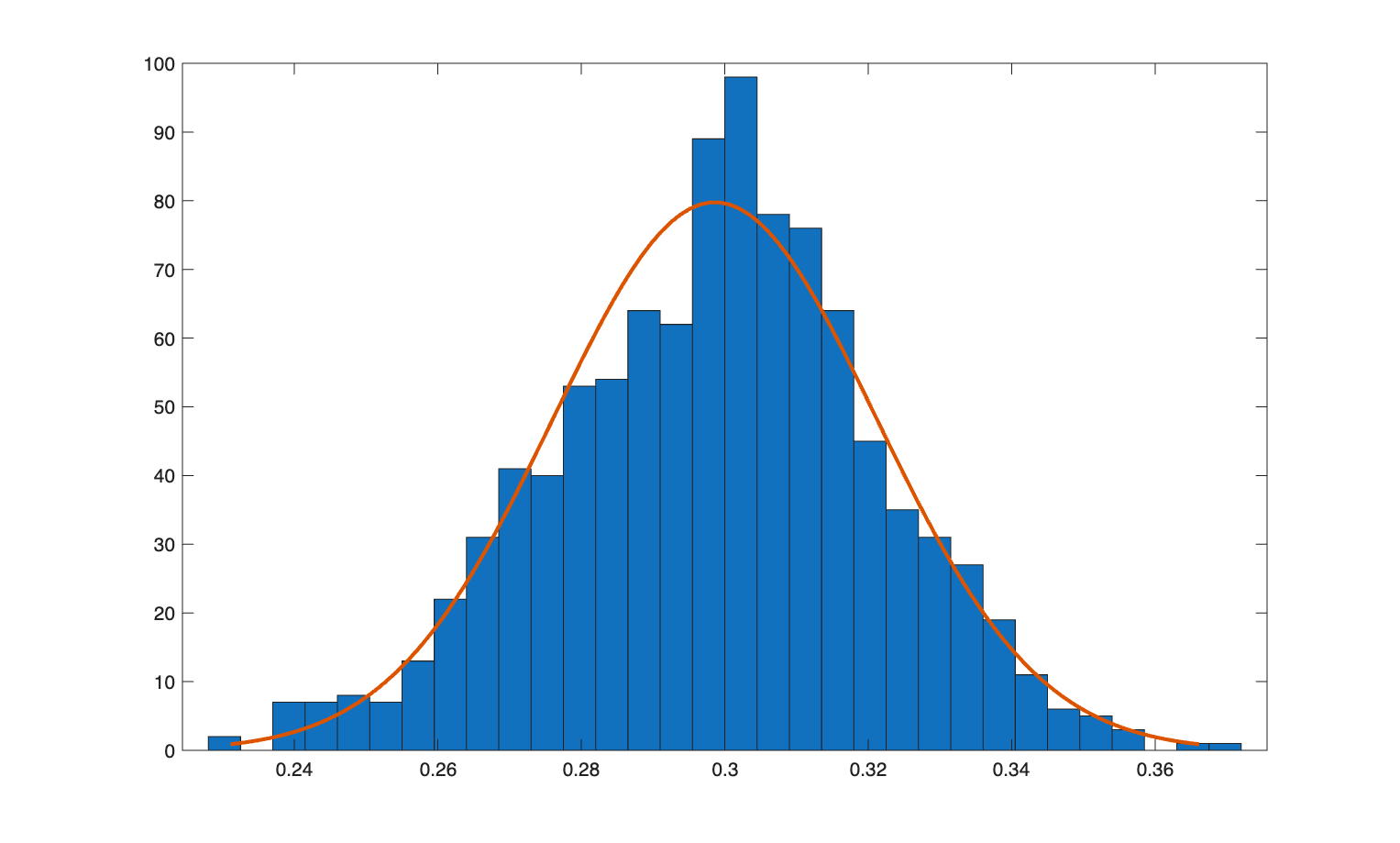}
        \caption{Histogram of $\hat{\psi_1}$ }

    \end{subfigure}
    \hfill
        \begin{subfigure}[b]{0.49\textwidth}
        \centering
        \includegraphics[width=\textwidth]{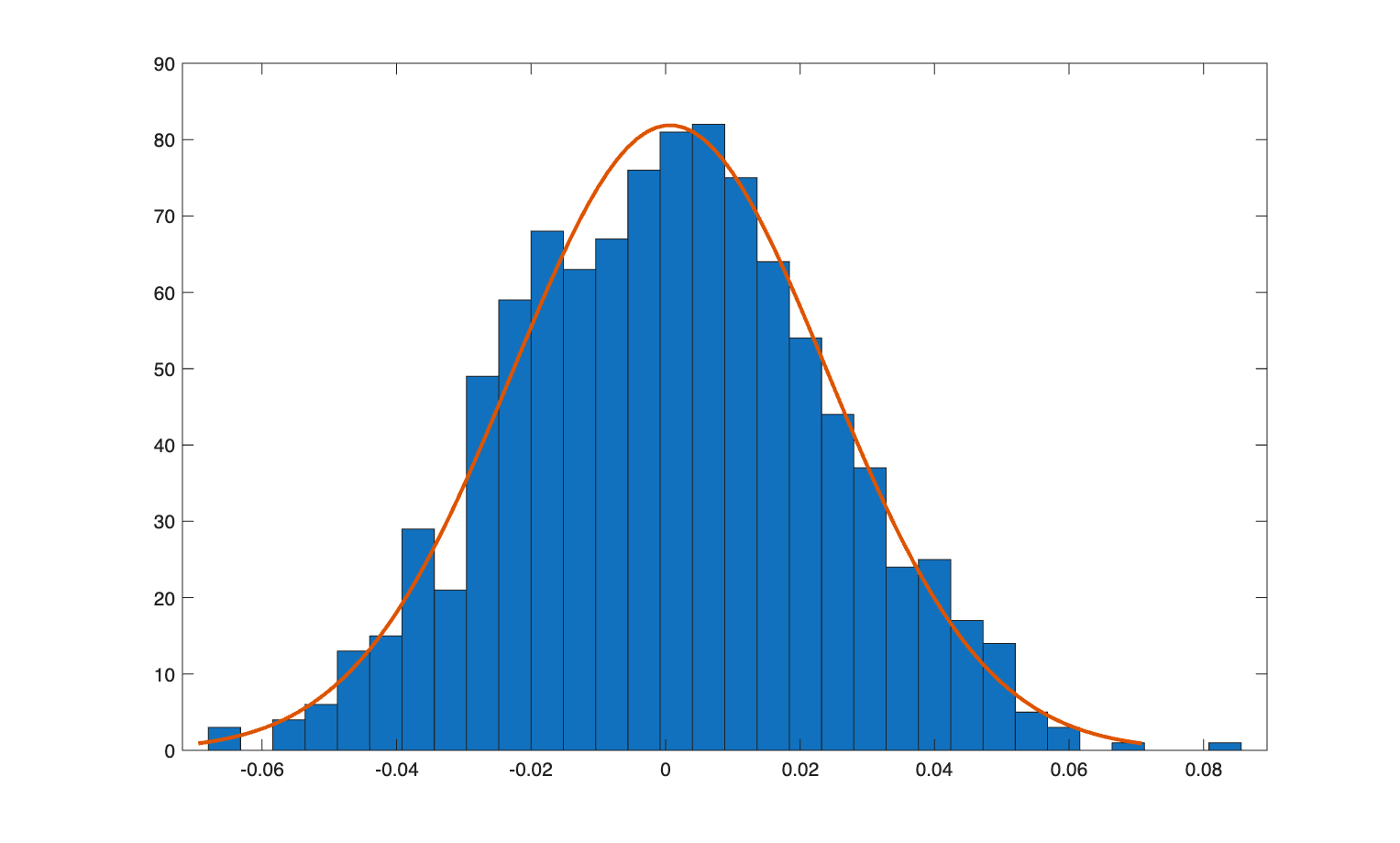}
        \caption{Histogram of $\hat{\psi_2}$ }

    \end{subfigure}
    \begin{subfigure}[b]{0.49\textwidth}
        \centering
        \includegraphics[width=\textwidth]{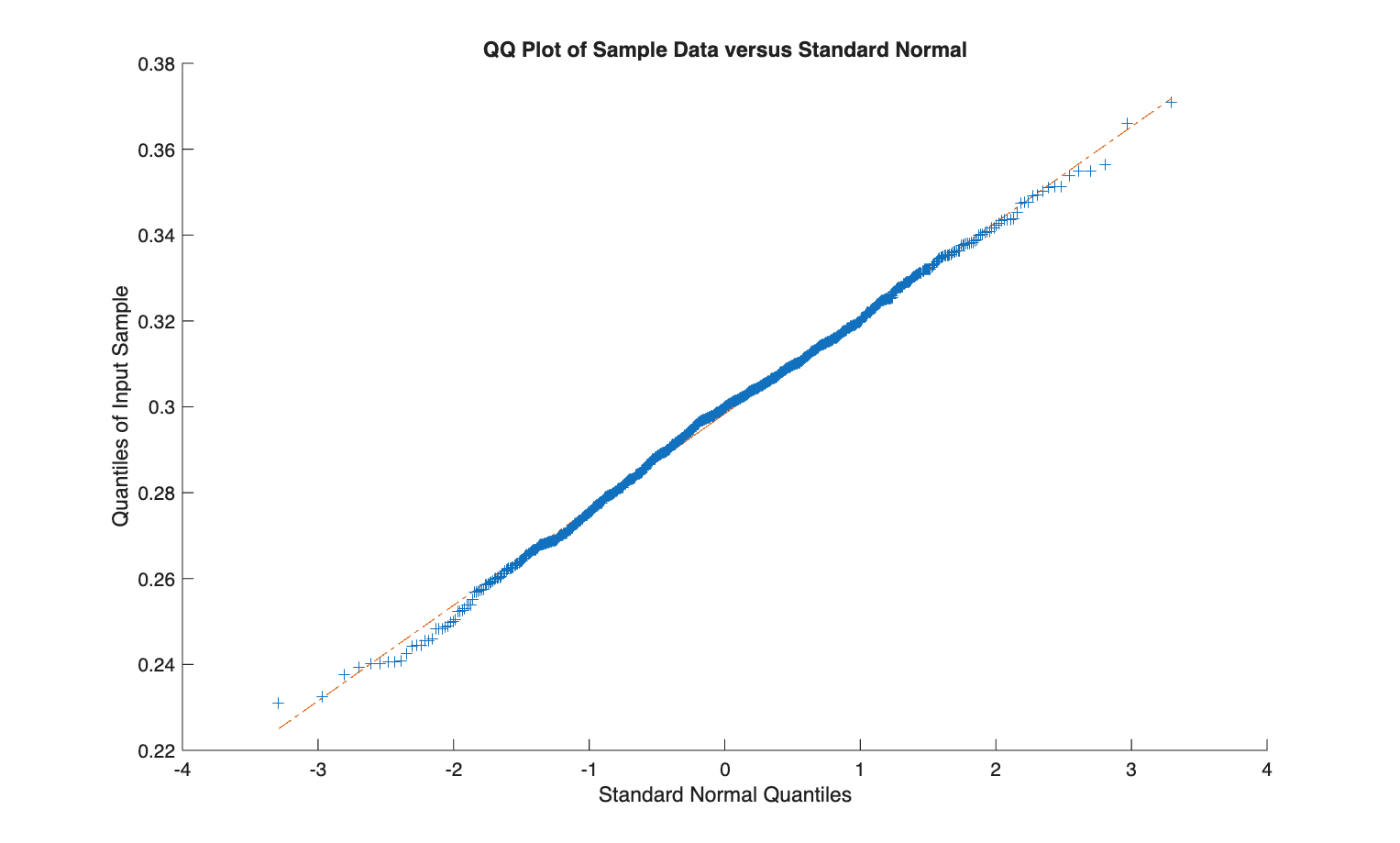}
        \caption{QQ-plot of $\hat{\psi_1}$ }
    \end{subfigure}
    \hfill
    \begin{subfigure}[b]{0.49\textwidth}
        \centering
        \includegraphics[width=\textwidth]{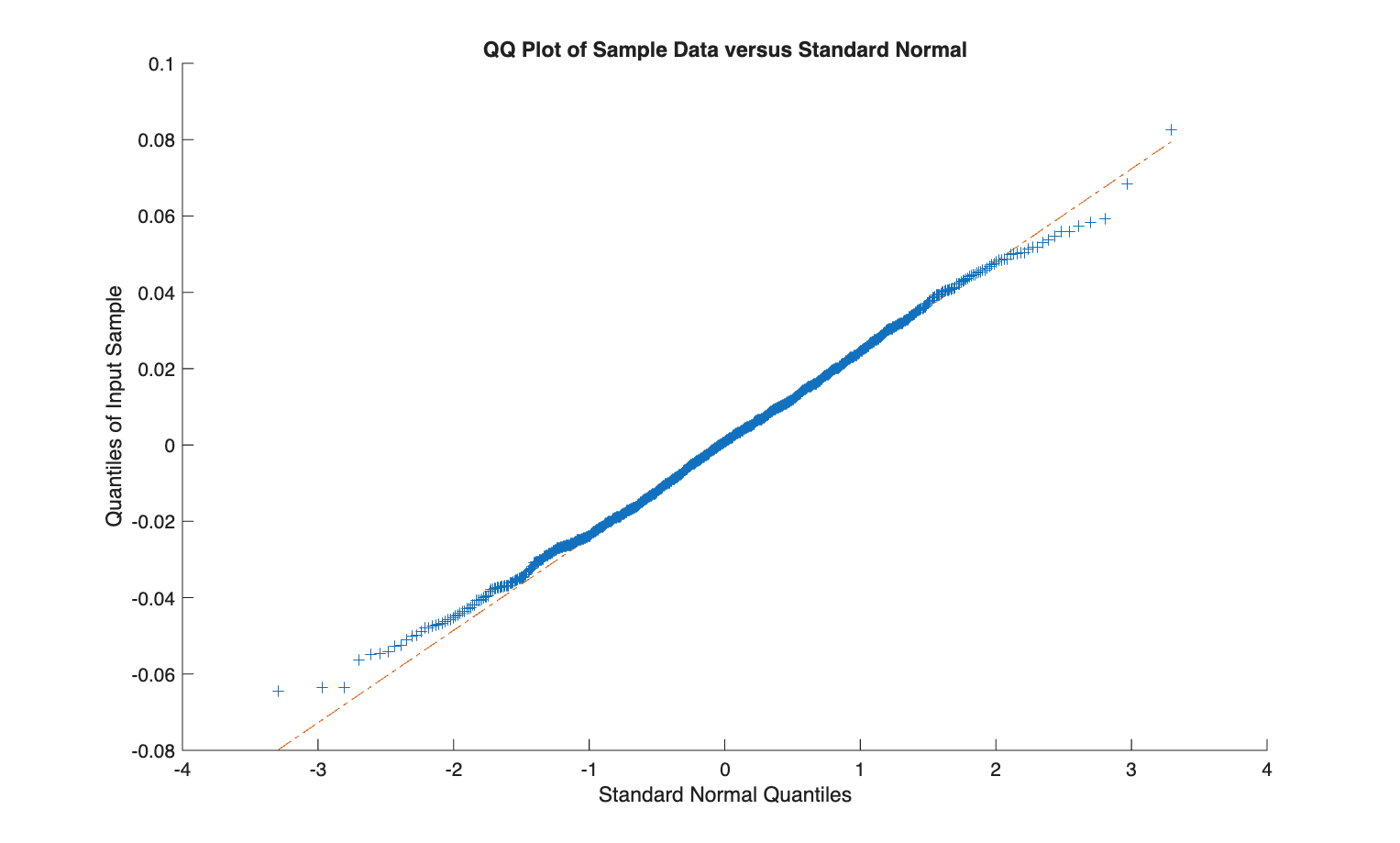}
        \caption{QQ-plot of $\hat{\psi_2}$}

    \end{subfigure}
    \caption{Histogram and QQ-plot of GCov estimates under misspecification (nested models)}
    \label{figure 3}
\end{figure}

To check the case under the non-nested scenario, we generate MAR(1,0) with $\phi=0.3$, $t(6)$ error distribution, $T=2000$ observations and then fit $MAR(0,2)$ which is constrained and we replicate 1000 times. Since we have the constrained version and the models are non-nested the close form of pseud-true value is not available. Therefore, we choose the average of $\hat{b}_{T,S}$ as the simulated  pseud-true value index and each $\hat{b}_{T,S}$ obtained from $S=100$ simulations.

 \begin{figure}[ht!]
    \centering
    
    \begin{subfigure}[b]{0.49\textwidth}
        \centering
        \includegraphics[width=\textwidth]{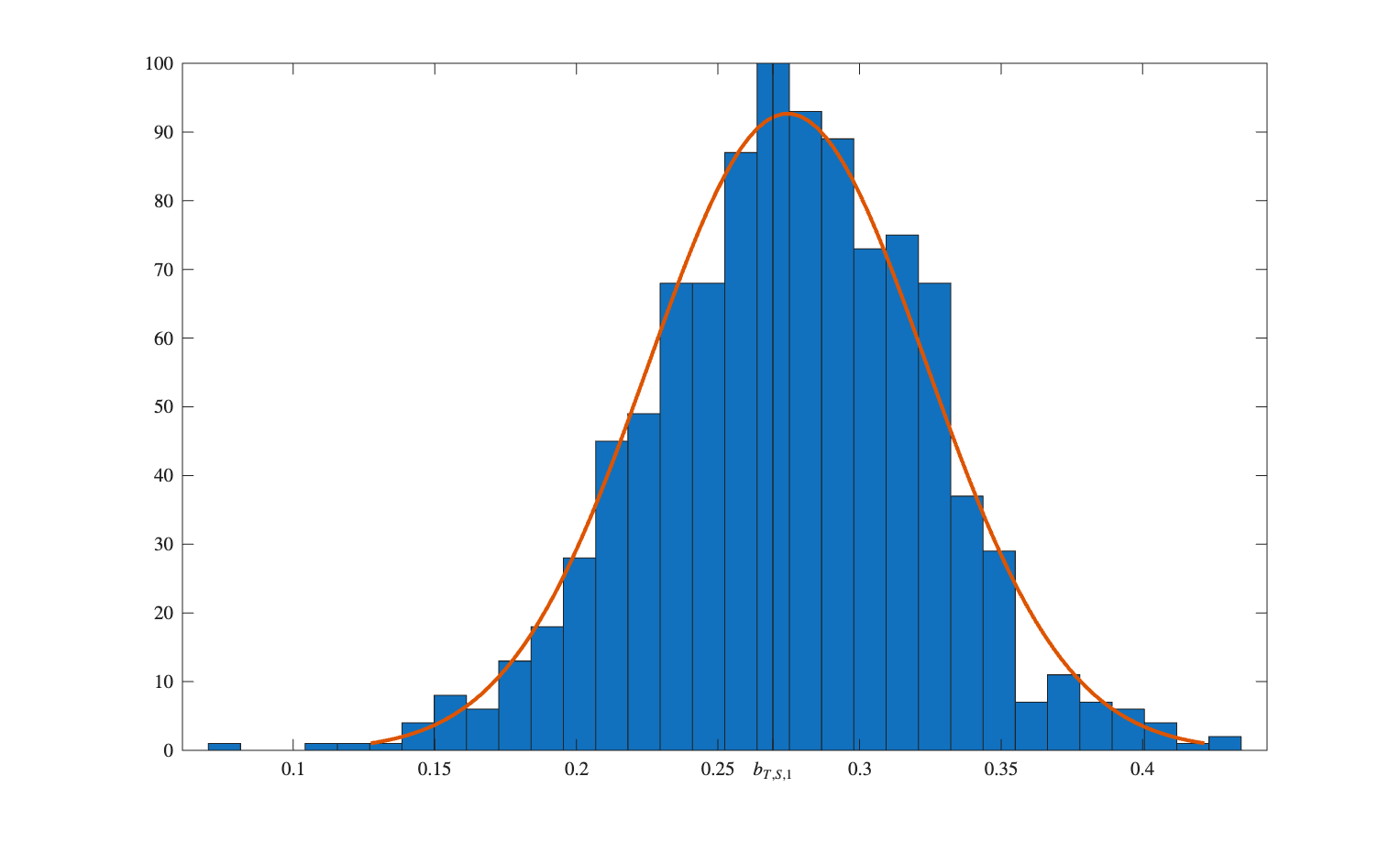}
        \caption{Histogram of $\hat{\psi_1}$ }

    \end{subfigure}
    \hfill
        \begin{subfigure}[b]{0.49\textwidth}
        \centering
        \includegraphics[width=\textwidth]{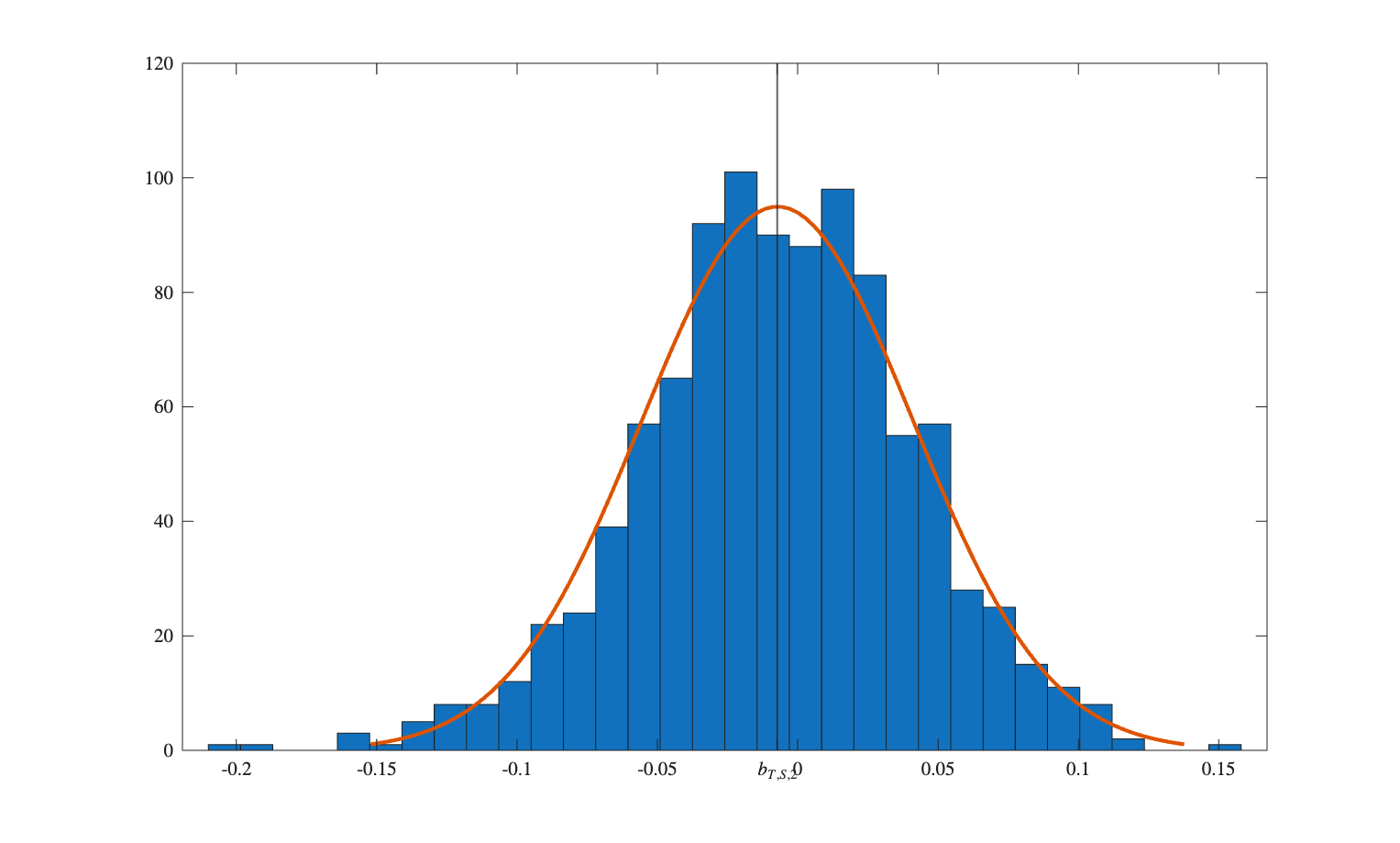}
        \caption{Histogram of $\hat{\psi_2}$ }

    \end{subfigure}
    \begin{subfigure}[b]{0.49\textwidth}
        \centering
        \includegraphics[width=\textwidth]{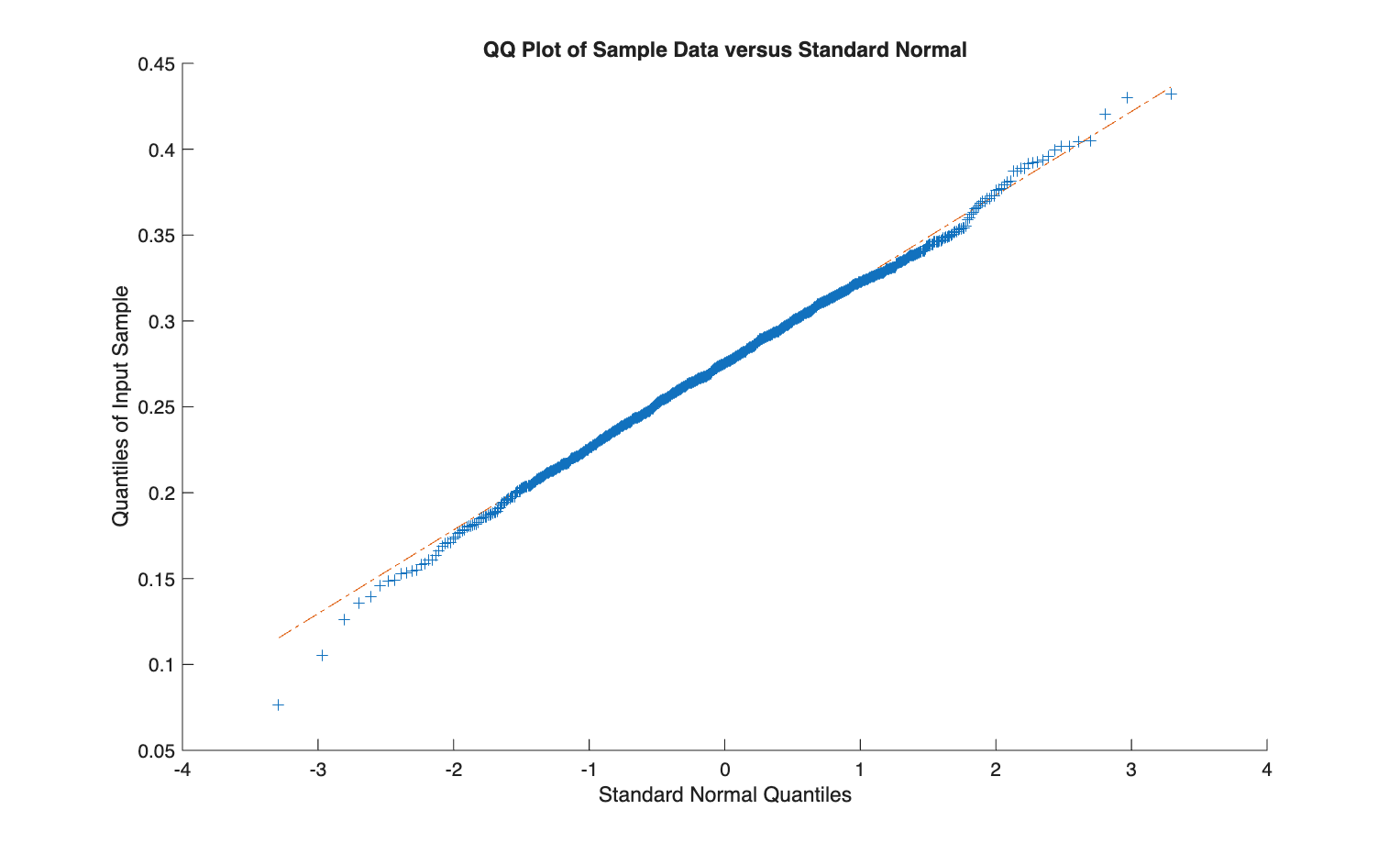}
        \caption{QQ-plot of $\hat{\psi_1}$ }
    \end{subfigure}
    \hfill
    \begin{subfigure}[b]{0.49\textwidth}
        \centering
        \includegraphics[width=\textwidth]{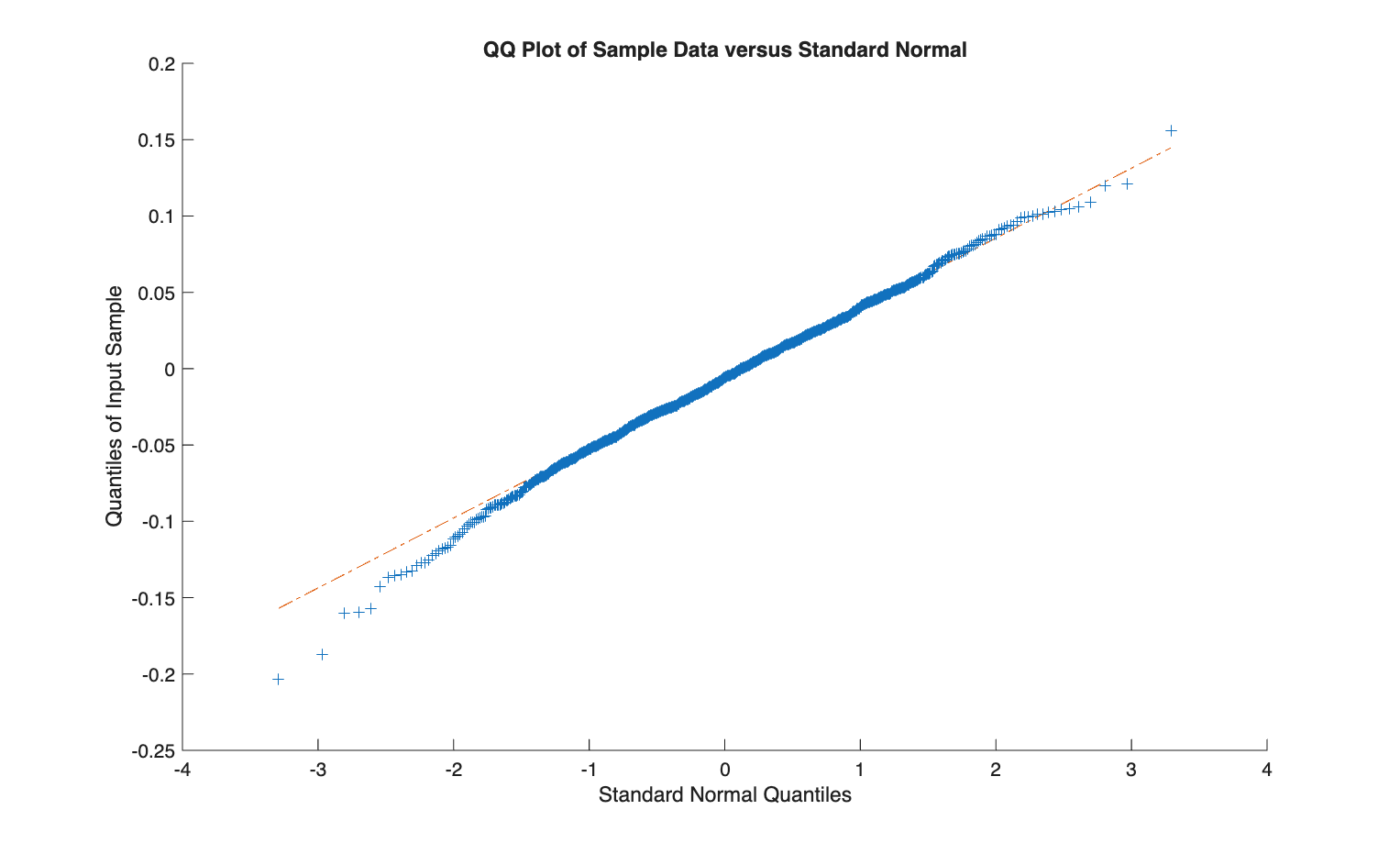}
        \caption{QQ-plot of $\hat{\psi_2}$}

    \end{subfigure}
    \caption{Histogram and QQ-plot of GCov estimates under misspecification (non-nested models)}
    \label{figure 4}
\end{figure}
\subsection{Test Simulations}
In the following examples, we use the GCov estimator with $K=2$ and $H=2$ and constrain the parameter space for each model. All DGPs have t(4), t(5), and t(6) error distributions, and we run a simulation experiment with $ T=500, 1000, 2000$ observations with 1000 replications.

\medskip
\nin \textbf{Nested Models.} Consider Null hypothesis of $M_1:MAR(0,1)$ and the alternative hypothesis is the  $M_2:MAR(0,2)$ model. Since we are in the nested test, we can use a Wald-type test with simplified $\Omega_A=J^a_{22}(b(\hat{\theta},\hat{f}))$. From Remark 6, we have
$$b_1 (\psi_{0,1})= \psi_{0,1}\; \text{and} \;  b_2 (\psi_{0,1})=0.$$
\nin Therefore we can construct the $\hat{\xi}^{W1}_T$ as follow 
$$\hat{\xi}^{W1}_T= T \begin{pmatrix}
    \hat{\psi_{2,1}}- \psi_{0,1} \\
    \hat{\psi_{2,2}}- 0
\end{pmatrix}' J^a_{22} \begin{pmatrix}
    \hat{\psi_{2,1}}- \psi_{0,1} \\
    \hat{\psi_{2,2}}- 0
\end{pmatrix}.$$
\nin Since the rank of $J^a_{22}$ is equal to 2 we compare the $\hat{\xi}^{W1}_T$ with $\chi^2_{0.95}(2)$. We reject the null hypothesis if $\hat{\xi}^{W1}_T>\chi^2_{0.95}(2)$. Table \ref{NESTEDSIMUAL} presents the empirical size and power of this test. 
\begin{table}[H]
\centering
\caption{ Empirical size and power of GCov-Based Wald test at 5\% significance level}

\begin{tabular}{cccccccccccc}
\hline
\multirow{2}{*}{S./P.} & \multirow{2}{*}{$\psi_1$} & \multirow{2}{*}{$\psi_2$} & \multicolumn{3}{c}{T=500}                                  & \multicolumn{3}{c}{T=1000}                              & \multicolumn{3}{c}{T=2000}                              \\ \cline{4-12} 
                       &                      &                      & \multicolumn{1}{c}{t(4)} & \multicolumn{1}{c}{t(5)}     & t(6)  & \multicolumn{1}{c}{t(4)} & \multicolumn{1}{c}{t(5)} & t(6)  & \multicolumn{1}{c}{t(4)} & \multicolumn{1}{c}{t(5)} & t(6)  \\ \hline
\multirow{2}{*}{S.}    &            0.3   &       &    0.071  &  0.066 &   0.071  &  0.069  &  0.066  &  0.070 &   0.058  &  0.063  &  0.053 \\  
                       &          0.7      &     & 0.078  &  0.083 &   0.089  &  0.066  &  0.063  &  0.066 &   0.058 &   0.058  &  0.055  \\ \hline

\multirow{2}{*}{P.}    & 0.7                  & -0.12                  & 0.711 &   0.720 &   0.708 &   0.932  &  0.929 &   0.926 &   0.995 &   0.998 &   0.998 \\ 
                       & 1.5                  & -0.56                  & 1                  & 1                      & 1  & 1                      & 1                      & 1     & 1                      & 1                      & 1     \\ \hline
\end{tabular}
\label{NESTEDSIMUAL}
\caption*{ S.: empirical size, P.: empirical power }
\end{table}

In the nested scenario, we can always construct $\hat{\xi}^{W1}_T$; however, since the pseudo-true values are not determined in the case of non-nested models when you constrain the parameter space of causal-noncausal models, we need to use $\hat{\xi}^{W3}_T$, which is based on simulated pseudo-true values. 

\medskip
\nin \textbf{Non-Nested Models}. Consider Null hypothesis of $M_1:MAR(1,0)$ and the alternative hypothesis is the  $M_2:MAR(0,2)$ model. Since we constrain the parameter space, these models are non-nested. Here we construct $\hat{\xi}^{W3}_T$ as indicated in \ref{GET3}. We use $S=100$ replications for the simulation algorithm of the test.  Table \ref{NONNESTEDSIMUAL} illustrates the empirical size and power of this test.

\begin{table}[H]
    \centering
    \caption{ Empirical size and power of Simulated GCov-Based Wald test at 5\% significance level}

\begin{tabular}{ccccccccccccc}
\hline
\multirow{2}{*}{S./P.} & \multirow{2}{*}{$\phi_1$}& \multirow{2}{*}{$\psi_1$} & \multirow{2}{*}{$\psi_2$} & \multicolumn{3}{c}{T=500}                                  & \multicolumn{3}{c}{T=1000}                              & \multicolumn{3}{c}{T=2000}                              \\ \cline{5-13} 
                       &                      &            &          & \multicolumn{1}{c}{t(4)} & \multicolumn{1}{c}{t(5)}     & t(6)  & \multicolumn{1}{c}{t(4)} & \multicolumn{1}{c}{t(5)} & t(6)  & \multicolumn{1}{c}{t(4)} & \multicolumn{1}{c}{t(5)} & t(6)  \\ \hline
                    \multirow{2}{*}{S.}    &  0.3   & &       & 0.028  &  0.019  &  0.022  &  0.026  &  0.026  &  0.019    & 0.027  &  0.026  &  0.016 \\
   &            0.7 &  &       &   0.032 &    0.020  &   0.025 &    0.028 &    0.022 &    0.026 &    0.027 &    0.018     &  0.011 \\ \hline

\multirow{2}{*}{P.}   & & 0.7                  & -0.12                  & 0.806 &    0.671 &    0.552 &    0.967 &    0.884 &    0.829 &    0.998 &    0.993 &    0.993  \\ 
                      & & 1.5                  & -0.56                  & 1                  & 1                      & 1  & 1                      & 1                      & 1     & 1                      & 1                      & 1     \\ \hline
\end{tabular}
\label{NONNESTEDSIMUAL}
\caption*{ S.: empirical size, P.: empirical power }
\end{table}

The empirical sizes in Table \ref{NONNESTEDSIMUAL} lie below the nominal 5\% level, so the simulated test $\hat{\xi}^{W3}_T$ is conservative in finite samples, in contrast to the mild oversizing of $\hat{\xi}^{W1}_T$ in the nested design. This behavior has a natural explanation. The statistic centers $\hat{\beta}_T$ at the simulated pseudo-true value $b_{T,S}(\hat{\theta},\hat{f})$, which is computed from artificial samples generated with the estimated parameter $\hat{\theta}$ and bootstrap draws of the fitted residuals of the observed series. Under the null hypothesis, sample-specific features of the data, in particular the realized configuration of extreme residual draws, which is influential under heavy-tailed $t$ innovations, are therefore transmitted to the simulated target, inducing a positive finite-sample correlation between $\hat{\beta}_T$ and $b_{T,S}(\hat{\theta},\hat{f})$. The variance $\Omega^a_S$ in Remark 2 captures the first-order effect of replacing $\theta_0$ by $\hat{\theta}$, but not this additional co-movement generated by resampling from the same residuals, so the dispersion of $\hat{\beta}_T-b_{T,S}(\hat{\theta},\hat{f})$ is overstated and the quadratic form is stochastically smaller than its $\chi^2$ limit. The constrained estimation reinforces this effect: both $\hat{\beta}_T$ and the simulated estimates $\hat{\beta}^s$ are subject to the same stationarity restrictions, which truncate their sampling distributions in the same direction and further compress their difference. Since the conservatism protects the maintained model against spurious rejection while leaving the power against distant alternatives essentially unaffected, we regard the size distortion as a benign feature of the simulated test, although refined critical values could be obtained by bootstrap methods.

\section{Empirical Application}

This section examines monthly data from the Producer Price Index by Commodity: Final Demand: Final Energy Demand (PPIDES) from November 2009 to January 2023. We detrend the series by regressing it on a constant and time. In the selection stage, we use both constrained and unconstrained GCov estimators to fit causal-noncausal processes, highlighting the importance of constrained GCov and the potential for misspecification under unconstrained GCov.  

First, by the NLSD test proposed by \cite{jasiak2023gcov}, we show the existence of linear and nonlinear dependence in the time series, considering $K=2$ transformations of residuals and squared residuals and $H=10$. The test value is 1245.1, and the 0.95 percent chi-square critical value is 55.76, indicating dependence in the time series. 

For models selection in this application we choose different sets of $MAR$ processes ander null and alternative. The structure we adopt consider the model with less total number of lag-leads as null hypothesis ($M_1$) and the model with more total number of lag-leads as alternative ($M_2$). Table \ref{modelselection} provides the results based on the simulated based test $\xi^{W_3}$. Based on Table \ref{modelselection}, we reject the $MAR(1,0)$ and $MAR(0,1)$ against $MAR(1,1)$ and we do not reject $MAR(1,1)$ against $MAR(2,1)$ and $MAR(1,2)$.

\begin{table}[ht!]
\centering
\caption{ Model selection based on $\hat{\xi}^{W3}_T$  }

\begin{tabular}{@{}clclcccccccc@{}}
\toprule
\multicolumn{2}{c}{\multirow{2}{*}{$M_1$}} & \multicolumn{2}{c}{\multirow{2}{*}{$M_2$}} & \multicolumn{2}{c}{$\hat{\theta}$} & \multicolumn{4}{c}{$\hat{\beta}$}                                 & \multirow{2}{*}{$\hat{\xi}^{W3}_T$ } & \multirow{2}{*}{$\chi^2_{0.95}$ } \\ \cmidrule(lr){5-10}
\multicolumn{2}{c}{}                      & \multicolumn{2}{c}{}                      & $\hat{\phi}$     & $\hat{\psi}$    & $\hat{\phi_1}$ & $\hat{\phi_2}$ & $\hat{\psi_1}$ & $\hat{\psi_2}$ &                       &                                  \\ \midrule
\multicolumn{2}{c}{MAR(1,0)}              & \multicolumn{2}{c}{\textbf{MAR(1,1)}}     & 0.99             &                 & 0.55           &                & 0.83           &                & 17.79                 & 5.99                             \\ \midrule
\multicolumn{2}{c}{MAR(0,1)}              & \multicolumn{2}{c}{\textbf{MAR(1,1)}}     &                  & 0.93            & 0.55           &                & 0.83           &                & 29.83                 & 5.99                             \\ \midrule
\multicolumn{2}{c}{\textbf{MAR(1,1)}}     & \multicolumn{2}{c}{MAR(2,1)}              & 0.55             & 0.83            & 0.58           & -0.11          & 0.85           &                & 1.25                  & 7.81                             \\ \midrule
\multicolumn{2}{c}{\textbf{MAR(1,1)}}     & \multicolumn{2}{c}{MAR(1,2)}              & 0.55             & 0.83            & 0.67           &                & 0.84           & -0.03          & 3.56                  & 7.81                             \\ \bottomrule
\end{tabular}

\label{modelselection}
\end{table}

\begin{table}[ht!]

\centering
\caption{ Estimated parameters of selected causal-noncausal models, GCov specification test with $\chi^2$ critical values at $5\%$ significance level, and roots of $\hat{\Phi}(L^{-1})$  and $\hat{\Psi}(L)$ }

\begin{tabular}{cccccccccc}
\hline
              Panel &   & $\phi_1$          & $\psi_1$  & $\psi_2$            & test statistic       & $\chi^2_{0.95}$          & $|L_1^{\phi}|>1$             & $|L_1^{\psi}|<1$             & $|L_2^{\psi}|<1$     \\ \hline
 \multirow{2}{*}{UC}&MAR(1,1) & 1.20$^*$ & 1.80$^*$ &   & 42.69 & 52.38 & 0.83 & 1.80 & \\ \cline{2-10}
&MAR(1,2)       & $1.25^*$   & $1.53^*$          & -0.07  & 27.14       & 52.19      & 0.79 & 0.05   & 1.49              \\ \hline

 \multirow{2}{*}{C}&\textbf{MAR(1,1)} & \textbf{0.55$^*$} & \textbf{0.83$^*$} &   & \textbf{42.69} & \textbf{52.38} & \textbf{1.80} & \textbf{0.83} & \textbf{} \\ \cline{2-10}
&MAR(1,2)       & $0.67^*$   & $0.85^*$          & -0.04  & 27.13        & 52.19      & 1.49 & 0.05   & 0.80               \\ \hline
\end{tabular}
\label{applications}
\caption*{* indicates statistical significance at 5\%}
\end{table}

\begin{figure}[ht!]
    \centering
    
    \begin{subfigure}[b]{0.49\textwidth}
        \centering
        \includegraphics[width=\textwidth]{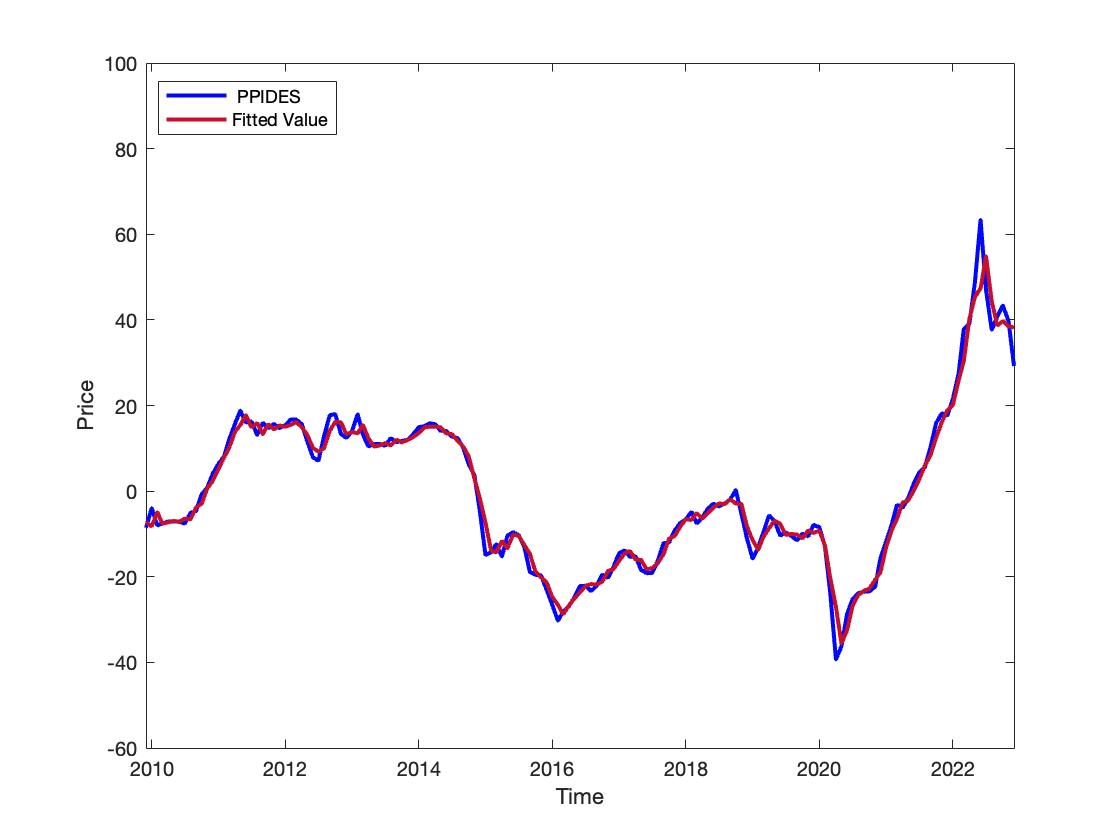}
        \caption{ PPIDES and fitted values of MAR(1,1) }
        \label{2-a}
    \end{subfigure}
    \hfill
        \begin{subfigure}[b]{0.49\textwidth}
        \centering
        \includegraphics[width=\textwidth]{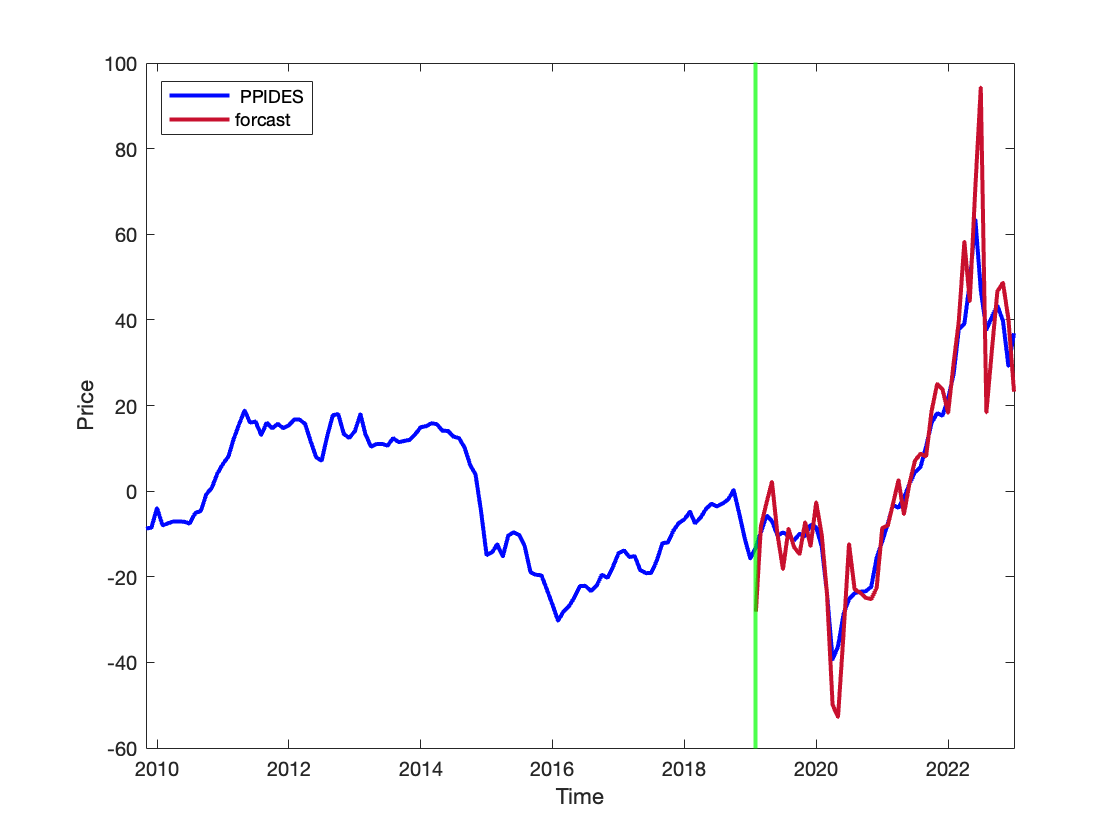}
        \caption{ One step a head forecast}
        \label{2-b}
    \end{subfigure}
    \caption{PPIDES, MAR(1,1) fitted values and residuals}
    \label{figure 5}
\end{figure}

\begin{figure}[]
    \centering
    
    \begin{subfigure}[b]{0.49\textwidth}
        \centering
        \includegraphics[width=\textwidth]{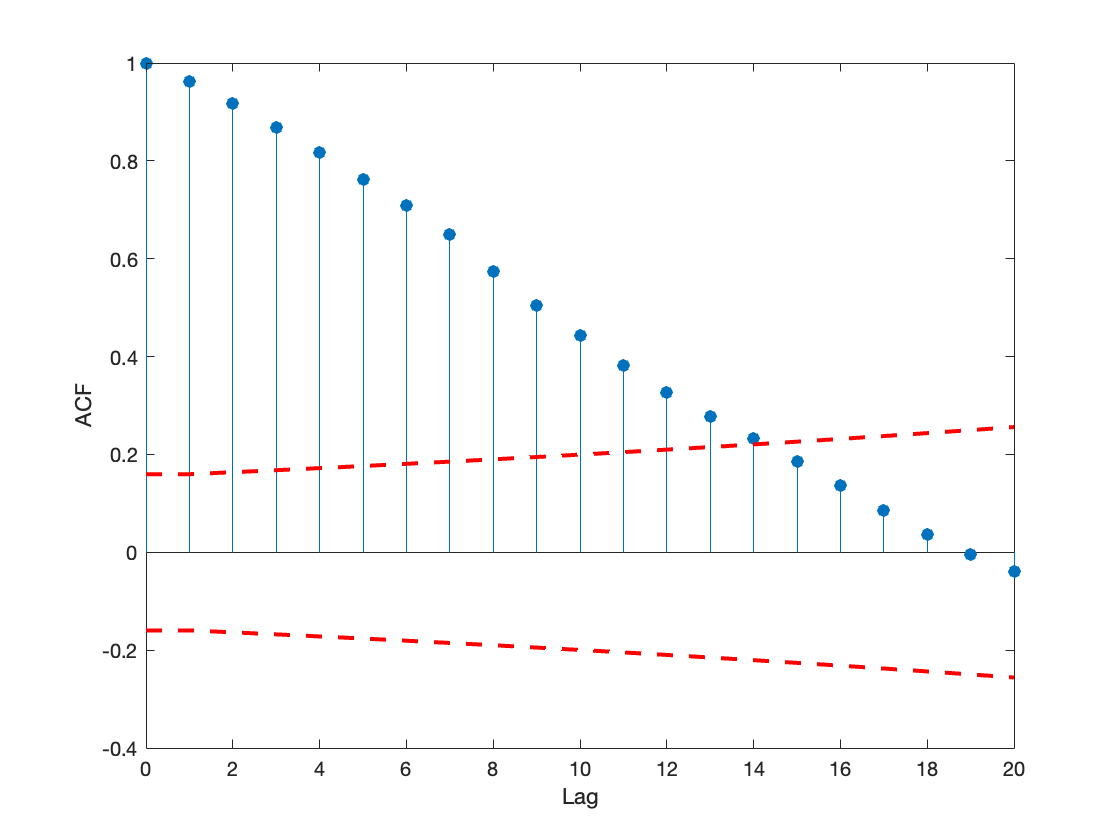}
        \caption{ACF of the series }
        \label{3-a}
    \end{subfigure}
    \hfill
        \begin{subfigure}[b]{0.49\textwidth}
        \centering
        \includegraphics[width=\textwidth]{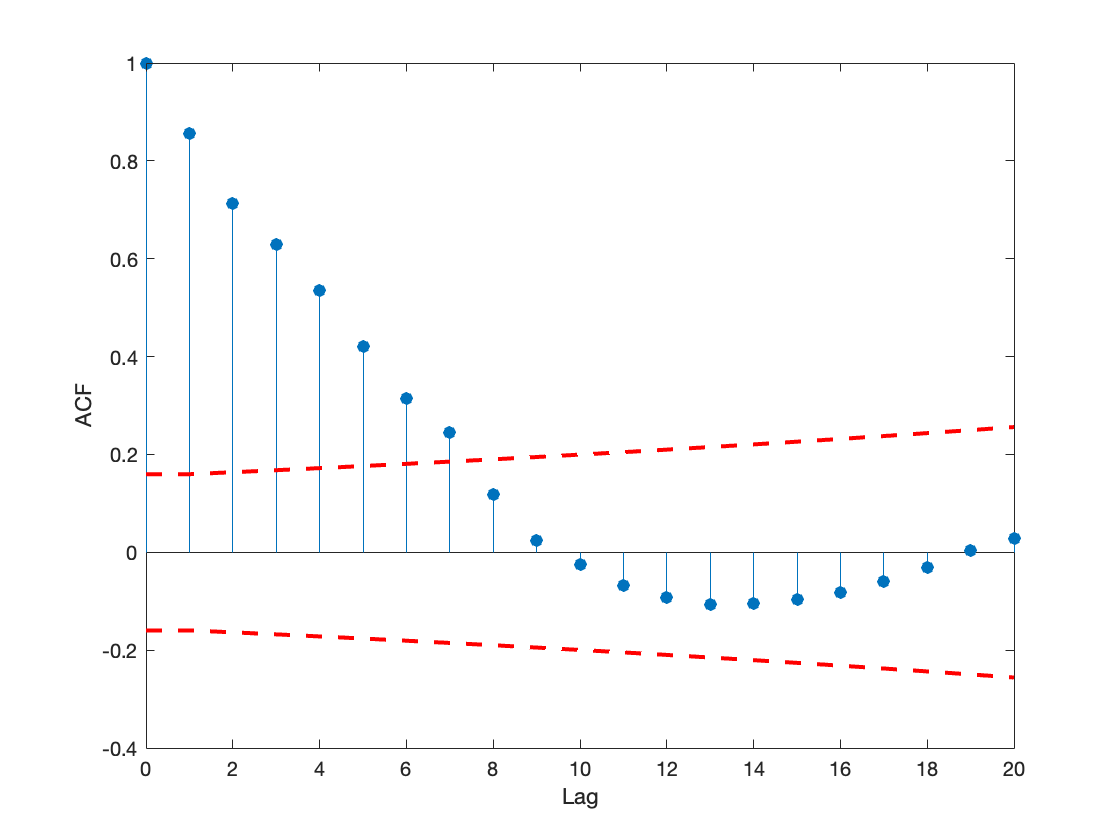}
        \caption{ACF of the series square}
        \label{3-b}
    \end{subfigure}
    \begin{subfigure}[b]{0.49\textwidth}
        \centering
        \includegraphics[width=\textwidth]{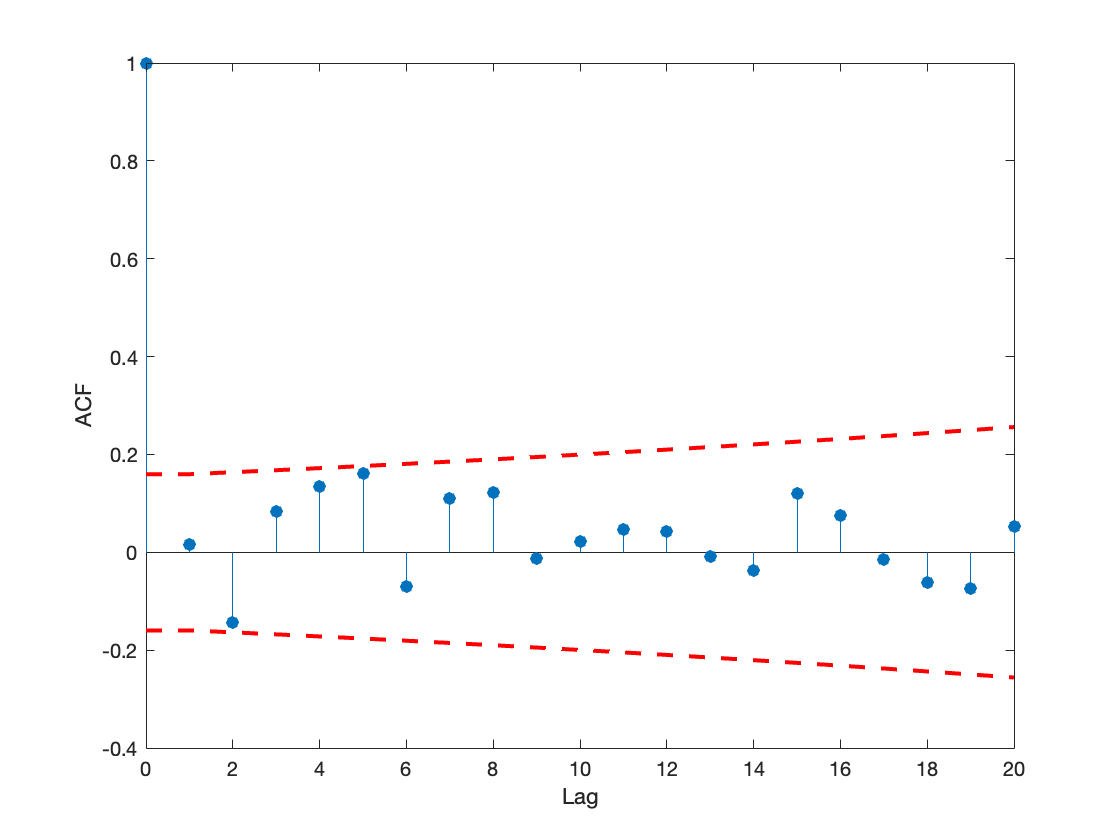}
        \caption{ACF of the residuals}
        \label{3-c}
    \end{subfigure}
    \hfill
    \begin{subfigure}[b]{0.49\textwidth}
        \centering
        \includegraphics[width=\textwidth]{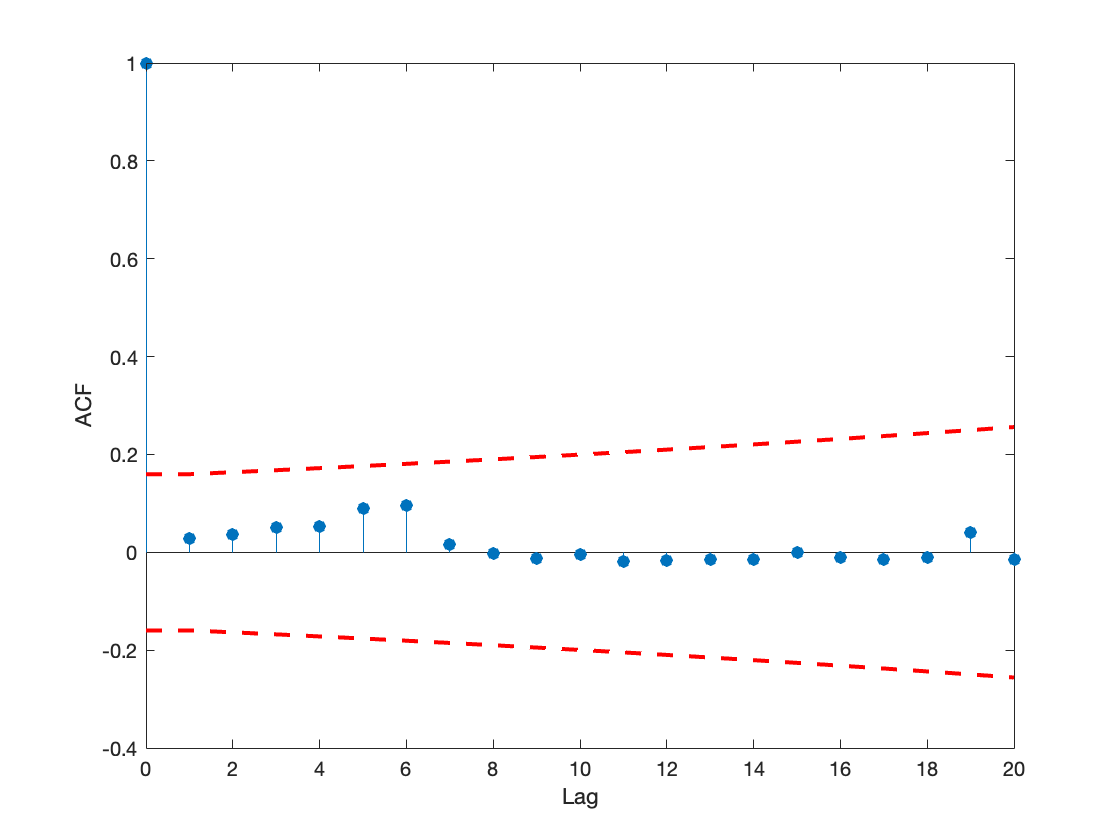}
        \caption{ACF of the square of the residuals}
        \label{3-d}
    \end{subfigure}
    \caption{ACF of series and fitted residuals by MAR(1,1)}
    \label{figure 6}
\end{figure}

Furthermore, by comparing the UC and C panels of Table \ref{applications}, we can argue that the unconstrained GCov provides pseudo-true values of the parameter, which is equal to the inverse of the true coefficients in this case, and violates the model's assumptions. However, the constrained one directly estimates the true specification. Figure \ref{figure 5} shows the fitted values of $MAR(1,1)$ (panel a) and the one-step-ahead forecast(panel b). Moreover, we report the ACF of the series, square series, residuals, and squared residuals in Figure \ref{figure 6} to support the correct specification of $MAR(1,1)$. Finally, we use a Wald-type test to exclude under-fitting possibilities, which, in this case, is reduced to the simple t-test. Since the $\psi_2$ is insignificant, we conclude that $MAR(1,1)$ is the best model for the PPIDES series.

\section{Conclusion}

\vspace{-0.2 cm}
This paper investigates the properties of the GCov estimator under misspecification. We propose Wald-type and score-type tests based on the GCov estimator for model selection and provide their asymptotic distribution. Specifically, we contribute to the literature on causal-noncausal processes by providing a broader range of estimation, model selection, and hypothesis-testing tools. 

This work can be extended in several ways. First, develop encompassing tests based on the GCov estimator that can choose between two misspecified models and recommend the one closest to the true specification. Second, we can extend the properties of the GCov to indirect inference for the estimation of noninvertible moving average models. Lastly, the asymptotic properties of the bootstrap GCov test when the parameter of interest lies on the boundary of the parameter space remain for future research.




\appendix
\setcounter{equation}{0}\def\theequation{A.\arabic{equation}}

\section{Appendix}

In this Appendix, we investigate the asymptotic distribution of the estimated covariance matrix and the GCov estimator around the parameter's pseudo-true value, the estimator's second-order expansion, and the properties of the GCov-based generalized Wald test.

\subsection{ Asymptotic Properties of Sample Covariance Under Misspecification}

\medskip

\nin \textbf{Lemma A.1:} The distribution of the estimated covariance matrix under the misspecified model is as follows
\begin{equation}
    \sqrt{T}vec\hat{\Gamma}^a(h;\hat{\beta})\sim N(\lambda(h),V_h),
\end{equation}
Where $\lambda(h)=\sqrt{T}vec(\Gamma^a(h,\hat{\beta}))$ and $V_h$ is variance. We can estimate $V_h$ by $\hat{V_h}$ as follow:

$$\hat{V}_h=\Delta_h(0)+\Sum_{i=1}^{T-1} w_{h,T} (\Delta_h(i)+\Delta_h(i)')$$

\nin where 

$$\Delta_h(k)= Cov(v_t(h),v_{t-k}(h)),$$

$$v_t(h)=vec(\hat{\epsilon}^a_t \hat{\epsilon}_{t-h}^{a'}).$$

\nin \textbf{Proof:} The proof follows the same structure of \cite{jasiak2023gcov} Appendix B, but we need to rewrite the LLN Regularity conditions to show the consistency of sample covariance estimator similar to the proposition B.1 of \cite{jasiak2023gcov} and similar to Proposition B.2 of \cite{jasiak2023gcov} we need to show the asymptotic normality and provide consistent estimator of the variance. 

\nin \textbf{- Regularity Conditions for LLN uniform in $\beta$}. 

\nin {\it 1. Conditions on the pseudo-true nonlinear dynamics}

 1.i) The observations satisfy the model:
 \begin{equation}
     h^*(\tilde{Y}_{T,t}; \beta_T) = v_t, 
 \end{equation}

 \nin where the $v_t$'s are i.i.d. with pdf $f_0$.

 1.ii) The function $h^*$ is invertible with respect to $Y_{T,t}$; then we can write:

 \begin{eqnarray}
  Y_{T,t} = h(v_t, Y_{T, t-1},...,Y_{T, t-p}; \beta_T).   
 \end{eqnarray}

1.iii) For each given $T$, ($Y_{T,t})$ with $t=1,...,T$, is a strictly stationary and ergodic solution of the autoregressive equation (A.3).

\nin {\it 2. Conditions on the parameters}

\nin Suppose that the parameter space is $B $, where $\beta \in B\subset \mathbf{R}^{dim(\beta)}$. We assume that:

2.i) $B$ is a compact set with non-empty interiors. 
 
2.ii) $b(\theta_0,f_0)$ the pseudo-true value of the parameter is in the interior of $B$.

\nin In particular, for $T$  sufficiently large, $\beta_T$ are in the interior of $B$.

\medskip

\nin {\it 3. Regularity conditions on the function $h^*$}

3.i) The functions $h_k^*(y ; \beta), \; k=1,...,K$ are continuously differentiable on the interior $B$.

3.ii) Let us define: $H_k^*(\tilde{y}) = Max_{(\beta) \in B} [g_k^* (\tilde{y}, \beta)]^2$. 
We assume $\mathbb{E}_0 H_k^*(\tilde{Y}) < \infty, \; k=1,...,K$ where $\mathbb{E}_0$ denotes the expectation computed for the process $(\tilde{Y}_t)$ associated with the "asymptotic" pseudo-true value of parameters $b(\theta_0,f_0)$.

3.iii(Uniform Lipschitz condition)

\nin There exists a measurable function $\mathcal{L}(\tilde{y})$ such that for all $\beta_1,\beta_2 \in B$,

$$||h_k^*(\tilde{y} ; \beta_1)-h_k^*(\tilde{y} ; \beta_2)||\leq \mathcal{L}(\tilde{y})||\beta_1-\beta_2||, \mathbb{E} \mathcal{L}(\tilde{Y_t})^2<\infty  $$

\medskip

\medskip
\nin {4. Near-Epoch Dependence (NED) under Misspecification}

\nin The transformed residuals and their products satisfy $L^2$--NED on the true
data-generating process $\{Y_{T,t}\}$.

\medskip
\nin For all $j,k=1,\dots,K$ and all $h\le H$, the processes
\[
h^*_j(\tilde Y_{T,t};\beta)
\quad \text{and} \quad
h^*_j(\tilde Y_{T,t};\beta)\,
h^*_k(\tilde Y_{T,t-h};\beta)
\]
are $L^2$--near-epoch dependent on $\{Y_{T,t}\}$, uniformly in $\beta\in B$, i.e.
\[
\sup_{\beta\in\mathcal{B}}
\mathbb{E}\Big[
h^*_j(\tilde Y_{T,t};\beta)
-
\mathbb{E}\big(
h^*_j(\tilde Y_{T,t};\beta)
\mid
Y_{T,t-r},\dots,Y_{T,t+r}
\big)
\Big]^2
\;\le\;
c_{T,t}\,\varphi(r),
\]
and
\[
\sup_{\beta\in B} 
\mathbb{E}\Big[
h^*_j(\tilde Y_{T,t};\beta)\,
h^*_k(\tilde Y_{T,t-h};\beta)
-
\mathbb{E}\big(
h^*_j(\tilde Y_{T,t};\beta)
\mid
Y_{T,t-r},\dots,Y_{T,t+r}
\big)
\mathbb{E}\big(
h^*_k(\tilde Y_{T,t-h};\beta)
\mid
Y_{T,t-r},\dots,Y_{T,t+r}
\big)
\Big]^2
\]
\[
\;\le\;
c_{T,t}\,\varphi(r).
\]

\medskip
\nin The function $\varphi(r)$ satisfies
\[
\varphi(r) \longrightarrow 0
\quad \text{as } r\to\infty,
\]
and
\[
\limsup_{T\to\infty}\frac{1}{T}\sum_{t=1}^T c_{T,t} < \infty.
\]

\medskip
\nin Under this condition, the transformed residuals and their autocovariance
products form a uniformly integrable mixingale array in the sense of
\cite{de1998weak}, which ensures the validity of the weak LLN of the sample
generalized autocovariances (\cite{andrews1988laws,andrews1992generic}).

\medskip
\nin \textbf{ - Regularity Conditions for the CLT under Misspecification}

\medskip
\nin Here we follow \cite{jasiak2023gcov} B.2 conditions, but instead of local alternatives we have them on a misspecified model. To establish the asymptotic normality of the sample generalized autocovariances
$\sqrt{T}\,\mathrm{vec}\,\hat{\Gamma}^a(h;\hat{\beta})$, we rely on a conditional
Lindeberg--Feller central limit theorem for multivariate martingale difference
triangular arrays (\cite{kundu2000central} and references in \cite{jasiak2023gcov} Appendix B.2).

\medskip
\nin Fix a lag $h\le H$ and let $\beta^\ast=b(\theta_0,f_0)$ denote the pseudo--true
parameter value.  Define
\[
v_t(h)
=
\mathrm{vec}\!\big(
h^*(\tilde Y_{T,t};\beta^\ast)\,
h^*(\tilde Y_{T,t-h};\beta^\ast)'
\big),
\]
and let ${\cal F}_{T,t}$ be the filtration generated by the triangular array
$\{Y_{T,\tau}:\tau\le t\}$.  Consider the martingale difference array
\[
X_{T,t}(h)
=
\frac{1}{\sqrt{T}}
\Big\{
v_t(h)
-
\mathbb{E}\big[v_t(h)\mid{\cal F}_{T,t-1}\big]
\Big\}.
\]

\medskip
\nin {\it (CLT.i) Martingale difference and finite second moments.}
The array $\{X_{T,t}(h),{\cal F}_{T,t}\}$ is a multivariate martingale difference
triangular array satisfying
\[
\mathbb{E}\big[X_{T,t}(h)\mid{\cal F}_{T,t-1}\big]=0,
\]
and each component of $X_{T,t}(h)$ has finite second-order moments, uniformly in $T$.

\medskip
\nin {\it (CLT.ii) Conditional variance convergence.}
For any conformable vector $b$, there exists a finite, positive semi-definite matrix
$V_h$ such that
\[
\sum_{t=1}^T
\mathbb{E}\!\left[
(b'X_{T,t}(h))^2
\mid{\cal F}_{T,t-1}
\right]
\stackrel{P}{\longrightarrow}
b'V_h b.
\]
The matrix $V_h$ coincides with the long-run covariance matrix of the vector process
$\{v_t(h)\}$.

\medskip
\nin {\it (CLT.iii) Conditional Lindeberg--Feller condition.}
For any $\epsilon>0$ and any conformable vector $b$,
\[
\sum_{t=1}^T
\mathbb{E}\!\left\{
(b'X_{T,t}(h))^2
\mathbf{1}_{\{|b'X_{T,t}(h)|>\epsilon\}}
\;\middle|\;
{\cal F}_{T,t-1}
\right\}
\stackrel{P}{\longrightarrow} 0.
\]

\medskip
\nin {\it (CLT.iv) Stability under triangular arrays.}
The moment bounds, NED conditions, and martingale approximations required for
{\rm (CLT.i)--(CLT.iii)} hold uniformly along the triangular array indexed by $T$,
so that the limiting distribution is stable under parameter sequences
$\beta_T\to\beta^\ast$.

\medskip
\nin Under conditions {\rm (CLT.i)--(CLT.iv)}, the conditional Lindeberg--Feller theorem
implies that
\[
\sum_{t=1}^T X_{T,t}(h)
\;\Rightarrow\;
N(0,V_h),
\]
which yields the asymptotic normality of the sample generalized autocovariances.

\medskip
\nin \textbf{ - Consistency of the HAC estimator of $V_h$.}

Define the centered sample version
\[
\bar v_T(h)=\frac{1}{T-h}\sum_{t=h+1}^{T} v_t(h),\qquad
\hat{\Delta}_h(k)=\frac{1}{T-h}\sum_{t=h+1+k}^{T}\big(v_t(h)-\bar v_T(h)\big)\big(v_{t-k}(h)-\bar v_T(h)\big)'.
\]
Let $w(\cdot)$ be a kernel with $w(0)=1$ and $w(x)=w(-x)$, and let the bandwidth satisfy
$L_T\to\infty$ and $L_T/T\to 0$. Define the HAC estimator
\[
\hat{V}_h=\hat{\Delta}_h(0)+\sum_{k=1}^{L_T} w\!\left(\frac{k}{L_T}\right)\Big(\hat{\Delta}_h(k)+\hat{\Delta}_h(k)'\Big).
\]
Under the $L^2$--NED on a strong mixing array and the kernel/bandwidth conditions
(so that $\{v_t(h)\}$ has covariance summability and the kernel estimator is consistent),
the kernel covariance consistency results imply
\[
\hat{V}_h\xrightarrow{p}V_h.
\]
Therefore, replacing $V_h$ by $\hat{V}_h$ yields feasible inference by Slutsky's theorem [\cite{de2000consistency}].
\hfill $\square$

\subsection{ First Order Condition}

From \cite{gourieroux2022generalized} supplementary material for $H=1$ we have:
\[
FOC= 2Tr \left(
\frac{\partial\hat{\Gamma}(1, \beta)}{\partial\beta_j}
\left[ \hat{\Gamma}(0, \beta)^{-1} \hat{\Gamma}(1, \beta)' \hat{\Gamma}(0, \beta)^{-1} \right]
\right)
\]
\[
- Tr \left\{
\left[ \hat{\tilde{R}}^2(1, \beta) \hat{\Gamma}(0, \beta)^{-1} + \hat{\Gamma}(0, \beta)^{-1}\hat{R}^2(1, \beta) \right]
\left[ \frac{\partial\hat{\Gamma}(0, \beta)}{\partial\beta} \right]
\right\},
\]
where
\[
\hat{\tilde{R}}^2(1; \beta) = \hat{\Gamma}(0; \beta)^{-1} \hat{\Gamma}(1; \beta)' \hat{\Gamma}(0; \beta)^{-1} \hat{\Gamma}(1; \beta),
\]
and
\[
\hat{R}^2(1; \beta) =  \hat{\Gamma}(1; \beta) \hat{\Gamma}(0; \beta)^{-1} \hat{\Gamma}(1; \beta)'\hat{\Gamma}(0; \beta)^{-1}
\]
for $j= 1,..,J$. However, here we rewrite it as:
\begin{equation}
    FOC= Tr[\hat{\Gamma}(0; \beta)^{-1}\hat{\Gamma}(1; \beta)' \hat{\Gamma}(0; \beta)^{-1}W(\beta)],
\end{equation}
where 
$$W(\beta)=\{ 2\frac{\partial\hat{\Gamma}(1; \beta)}{\partial\beta}- \hat{\Gamma}(1; \beta) \hat{\Gamma}(0; \beta)^{-1}\frac{\partial\hat{\Gamma}(0; \beta)}{\partial\beta_j} - \frac{\partial\hat{\Gamma}(0; \beta)}{\partial\beta_j} \hat{\Gamma}(0; \beta)^{-1}  \hat{\Gamma}(1; \beta)\}.$$

\medskip

\subsection{ Second Order Asymptotic Expansion}
\[
d_{FOCj}(\beta) = 2Tr \left(
\frac{\partial\hat{\Gamma}(1, \beta)}{\partial\beta_j}
d\left[ \hat{\Gamma}(0, \beta)^{-1} \hat{\Gamma}(1, \beta)' \hat{\Gamma}(0, \beta)^{-1} \right]
\right)
\]

\[
+ 2Tr \left( \hat{\Gamma}(0, \beta)^{-1} \hat{\Gamma}(1, \beta) \hat{\Gamma}(0, \beta)^{-1}
d\left[ \frac{\partial\hat{\Gamma}(1, \beta)}{\partial\beta_j} \right] \right)
\]

\[
- Tr \left\{
\left[ \hat{\tilde{R}}^2(1, \beta) \hat{\Gamma}(0, \beta)^{-1} - \hat{\Gamma}(0, \beta)^{-1}\hat{R}^2(1, \beta) \right]
d\left[ \frac{\partial\hat{\Gamma}(0, \beta)}{\partial\beta_j} \right]
\right\}
\]

\[
- Tr \left\{
\left[
\frac{\partial\hat{\Gamma}(0, \beta)}{\partial\beta_j}
\right]
d\left[ \hat{\tilde{R}}^2(1, \beta) \hat{\Gamma}(0, \beta)^{-1} - \hat{\Gamma}(0, \beta)^{-1} \hat{R}^2(1, \beta) \right]
\right\}.
\]
\medskip These are the results from \cite{gourieroux2022generalized} supplementary material. However, we can rewrite it as:
\[
d_{FOCj}(\beta) = Tr\left\{\hat{\Gamma}(0; \beta)^{-1} d[ \hat{\Gamma}(1; \beta)'] \hat{\Gamma}(0; \beta)^{-1} W(\beta) + \hat{\Gamma}(0; \beta)^{-1}  \hat{\Gamma}(1; \beta)' \hat{\Gamma}(0; \beta)^{-1} d W(\beta) \right\}.
\]

Based on $d(A(\beta)+B(\beta))=d(A(\beta))+d(B(\beta))$ and $d(A(\beta)B(\beta))=d(A(\beta))B(\beta)+A(\beta)d(B(\beta))$ we have:
\[
dW(\beta)=\{ 2d[\frac{\partial\hat{\Gamma}(1; \beta)}{\partial\beta_j}]- d(\hat{\Gamma}(1; \beta)+ \hat{\Gamma}(1; \beta)') \hat{\Gamma}(0; \beta)^{-1}\frac{\partial\hat{\Gamma}(0; \beta)^{-1}}{\partial\beta_j}
\]
\[
-(\hat{\Gamma}(1; \beta)+ \hat{\Gamma}(1; \beta)') d[\hat{\Gamma}(0; \beta)^{-1}\frac{\partial\hat{\Gamma}(0; \beta)^{-1}}{\partial\beta_j}] \}.
\]
\[
=\{ 2d[\frac{\partial\hat{\Gamma}(1; \beta)}{\partial\beta_j}]- (d\hat{\Gamma}(1; \beta)+ d\hat{\Gamma}(1; \beta)') \hat{\Gamma}(0; \beta)^{-1}\frac{\partial\hat{\Gamma}(0; \beta)^{-1}}{\partial\beta_j}
\]
\[
-(\hat{\Gamma}(1; \beta)+ \hat{\Gamma}(1; \beta)') [d(\hat{\Gamma}(0; \beta)^{-1})\frac{\partial\hat{\Gamma}(0; \beta)^{-1}}{\partial\beta_j}+ \hat{\Gamma}(0; \beta)^{-1}d(\frac{\partial\hat{\Gamma}(0; \beta)^{-1}}{\partial\beta_j})] \}.
\]
Therefore the matrix $J(h,b(\theta_0,f_0))$ has elements $(j,k)$ as follow
\[
-J^*(h,b(\theta_0,f_0))=Tr\left\{\hat{\Gamma}(0; \beta)^{-1}  \frac{\partial\hat{\Gamma}(h; \beta)'}{\partial\beta_k} \hat{\Gamma}(0; \beta)^{-1} W(h,\beta) + \hat{\Gamma}(0; \beta)^{-1}  \hat{\Gamma}(h; \beta)' \hat{\Gamma}(0; \beta)^{-1} \frac{W(h,\beta)}{\partial\beta_k} \right\}
\]
where
$$W(h,\beta)=\{ 2\frac{\partial\hat{\Gamma}(h; \beta)}{\partial\beta_j}- \hat{\Gamma}(h; \beta) \hat{\Gamma}(0; \beta)^{-1}\frac{\partial\hat{\Gamma}(0; \beta)^{-1}}{\partial\beta_j} - \hat{\Gamma}(h; \beta)' \hat{\Gamma}(0; \beta)^{-1}\frac{\partial\hat{\Gamma}(0; \beta)^{-1}}{\partial\beta_j}\},$$
and
\[
\frac{W(h,\beta)}{\partial\beta_k}=\{ 2\frac{\partial^2\hat{\Gamma}(h; \beta)}{\partial\beta_j\partial\beta_k}- (\frac{\partial\hat{\Gamma}(h; \beta)}{\partial\beta_k}+ \frac{\partial\hat{\Gamma}(h; \beta)'}{\partial\beta_k}) \hat{\Gamma}(0; \beta)^{-1}\frac{\partial\hat{\Gamma}(0; \beta)^{-1}}{\partial\beta_j}
\]
\[
-(\hat{\Gamma}(h; \beta)+ \hat{\Gamma}(h; \beta)') [\frac{\partial\hat{\Gamma}(0; \beta)^{-1}}{\partial\beta_k} 
 \frac{\partial\hat{\Gamma}(0; \beta)^{-1}}{\partial\beta_j}+ \hat{\Gamma}(0; \beta)^{-1}\frac{\partial^2\hat{\Gamma}(0; \beta)^{-1}}{\partial\beta_j \partial\beta_k}]\}.
\]
moreover, by $TR(AB)=Tr(BA)$ and $vec(ABC)=(C'\otimes A)vecB$ [See \cite{gourieroux2022generalized} Appendix 1 in supplementary material and their references] we have
\[
-J^*(h,b(\theta_0,f_0))=vec(W(h,\beta)[\hat{\Gamma}(0; \beta)^{-1} \otimes \hat{\Gamma}(0; \beta)^{-1} ]vec(\frac{\partial \hat{\Gamma}(h; \beta)}{\partial \beta})
\]
\[+ vec(\frac{W(h,\beta)}{\partial \beta})[\hat{\Gamma}(0; \beta)^{-1} \otimes \hat{\Gamma}(0; \beta)^{-1} ] vec(\hat{\Gamma}(h; \beta)).
\]

\subsection{ Proof of Proposition 1}

For $H=1$, we have $\sqrt{T}(\hat{\beta}-b(\theta_0,f_0))=J(b(\theta_0,f_0))^{-1}\sqrt{T}X(\hat{\Gamma})+O_P(1)$ following the same approach as \cite{gourieroux2022generalized} where $X(\hat{\Gamma})$ defined as:
$$
X^*(\hat{\Gamma}) =Tr[ \hat{\Gamma}(0; \beta)^{-1}\hat{\Gamma}(1; \beta)' \hat{\Gamma}(0; \beta)^{-1}W(\beta)].
$$
By the law of large number and Central Limit Theorem $\hat{\Gamma}$ goes to $\Gamma$ Asymptotically. 

\medskip Based on $TR(AB)=Tr(BA)$ we can rewrite $X(\hat{\Gamma})$ as
$$X^*(\hat{\Gamma}) = Tr[ \hat{\Gamma}(0; \beta)^{-1}W(\beta) \hat{\Gamma}(0; \beta)^{-1} \hat{\Gamma}(1; \beta)'],$$
and by $Tr(AB)=vec(A)vec(B)$ we have:
$$\sqrt{T}X^*(\hat{\Gamma}) = vec( \hat{\Gamma}(0; \beta)^{-1}W(\beta) \hat{\Gamma}(0; \beta)^{-1})vec(\hat{\Gamma}(1; \beta)')=A(\beta)vec(\sqrt{T}\hat{\Gamma}(1; \beta)')$$ 
By using the equality $vec(ABC)=(C'\otimes A)vecB$, we have:
$$A^*(\beta)= vec( \hat{\Gamma}(0; \beta)^{-1}W(\beta) \hat{\Gamma}(0; \beta)^{-1}) = vec(W(\beta))[\hat{\Gamma}(0; \beta)^{-1} \otimes \hat{\Gamma}(0; \beta)^{-1}].$$
 From Lemma A.1, when we are not under null, we have
 $$vec(\sqrt{T}\hat{\Gamma}(1; \beta)') \sim N(\lambda,\hat{\Sigma} \otimes \hat{\Gamma}(0; \beta)).$$
 The direct conclusion of this distribution is $\sqrt{T}X(\hat{\Gamma})$ has asymptotically normal distribution with variance
  $$I^*(1,b(\theta_0,f_0))=V_{asy}[\sqrt{T}X(\hat{\Gamma})]= vec(W(h,\beta))[\hat{\Gamma}(0; \beta)^{-1}   \otimes \hat{\Gamma}(0; \beta)^{-1}  ] \hat{V_1} [\hat{\Gamma}(0; \beta)^{-1}   \otimes \hat{\Gamma}(0; \beta)^{-1}  ] \vec(W(h,\beta))' .$$
 Therefore, $\sqrt{T}(\hat{\beta}-b(\theta_0,f_0))$ has normal distribution by mean of $\lambda(1)$ and asymptotic variance equal to:
$$ \Omega^*(1,b(\theta_0,f_0))=J^*(1,b(\theta_0,f_0))^{-1}I^*(1,b(\theta_0,f_0))J^*(1,b(\theta_0,f_0))^{-1}$$
for H=1.

\medskip
\subsection{ Proof of Corollary 1} 

From Proposition 1 and the asymptotic normality of the GCov estimator under correct specification. For a more detailed proof, see \cite{gourieroux1983testing} Appendix 1. 

\medskip
\subsection{ Proof of Proposition 2}

\medskip
\nin For $b(\hat{\theta},\hat{f})$ we have:

$$\sqrt{T}(\hat{\beta}-b(\hat{\theta},\hat{f}))= \sqrt{T}(\hat{\beta}-b(\theta_0,f_0))-\sqrt{T}(b(\hat{\theta},\hat{f}) -b(\theta_0,f_0)).$$

 The estimation of $b_T(\hat{\theta},\hat{f})$ and $b(\hat{\theta},\hat{f})$ can be done by minimizing the Kullback information criteria, which are based on the conditional distribution and first-order conditions  as follows 

\begin{equation}
    \sum_{h=1}^{H} \frac{\partial \mathrm{Tr}[\hat{R}_a^2(h, \beta)] }{\partial \beta } [b_T(\theta_0)] =0,
\end{equation}

\nin and

\begin{equation}
    \sum_{h=1}^{H} \frac{\partial \mathrm{Tr}[R_a^2(h, \beta)] }{\partial \beta } [b(\theta_0,f_0)] =0.
\end{equation}

\nin The asymptotic distribution of the $\sqrt{T}(b(\hat{\theta},\hat{f})-b(\theta_0,f_0))$ can be sustained by its equivalent $\frac{\partial b(\theta_0,f_0)}{\partial \theta'} \sqrt{T}(\hat{\theta}-\theta_0)$. Then by differentiation of equation 9 we can substitute term $\frac{\partial b(\theta_0,f_0)}{\partial \theta'}$ by ${J^a_{22}}^{-1}I^a_{21}$[differentiation of A.4 and A.5 with respect to $\theta_0$].Then we know $\sqrt{T}(b(\hat{\theta},\hat{f}) -b(\theta_0,f_0))$ is equivalent to ${J^a_{22}}^{-1}I^a_{21} (\hat{\theta}- \theta_0)$ and the asymptotic distribution fo the $\sqrt{T}(\hat{\beta}-b(\theta_0,f_0))$ comes from Propostion 2.2 and Corollary 1. Therefore

$$\sqrt{T}(\hat{\beta}-b(\hat{\theta},\hat{f}))= ({J^a_{22}}^{-1}I^a_{21}, I) \begin{pmatrix}
    \hat{\theta}_T - \theta_0 \\
    \hat{\beta}_T- b(\theta_0,f_0)
\end{pmatrix} +o_p(1),$$

\nin which indicates that 

$$\Omega^a_{A} = {J^a_{22}}^{-1} [ I^a_{22}  - I^a_{21}  {I^a_{11}} ^{-1} I^a_{12}] {J^a_{22}}^{-1}.$$


\medskip
\subsection{Proof of Corollary 2}

It is a direct consequence of the asymptotic normal distribution of the estimators provided in Proposition 1, Corollary 1, and Proposition 2.

\medskip
\subsection{Proof of Proposition 3}

Let us write the expansion of $\hat{\lambda}_T^{(1)}[b(\hat{\theta},\hat{f})]$ around $b(\theta_0,f_0)$ 
\begin{eqnarray*}
    \sqrt{T}\hat{\lambda}_T^{(1)}  &=& \sum_{h=1}^{H} \frac{\partial \mathrm{Tr}[R_a^2(h, \beta)] }{\partial \beta } [b(\hat{\theta},\hat{f})] \\
     & =& \sum_{h=1}^{H} \frac{\partial \mathrm{Tr}[R_a^2(h, \beta)] }{\partial \beta } [b(\theta_0,f_0)] \\
      & + & J^a_{22}[b(\theta_0,f_0)] (b(\hat{\theta},\hat{f})-b(\theta_0,f_0)) +o_p(1), 
\end{eqnarray*}

\nin and from the expansion of $\sum_{h=1}^{H} \frac{\partial \mathrm{Tr}[R_a^2(h, \beta)] }{\partial \beta } [\hat{\beta}]$ we have

\begin{eqnarray*}
    0 & = & \sum_{h=1}^{H} \frac{\partial \mathrm{Tr}[R_a^2(h, \beta)] }{\partial \beta } [\hat{\beta}] \\
     & = & \sum_{h=1}^{H} \frac{\partial \mathrm{Tr}[R_a^2(h, \beta)] }{\partial \beta } [b(\theta_0,f_0)] \\
      & + & J^a_{22}[b(\theta_0,f_0)] (\hat{\beta}-b(\theta_0,f_0)) +o_p(1). 
\end{eqnarray*}

\nin Therefore, $\sqrt{T}\hat{\lambda}_T^{(1)}$ is qual to $J^a_{22}\sqrt{T}(b(\hat{\theta},\hat{f})-\hat{\beta})$ which is asymptotically normal with variance equal to $I^{a}_{22}- I^a_{21} {I^a_{11}}^{-1} I^a_{12}$. The asymptotic distribution of the $\sqrt{T}\hat{\lambda}_T^{(2)}$ is similar.

\medskip
\subsection{Proof of Corollary 3}

A direct consequence of Proposition 3.

\bibliography{bibliography.bib}

\end{document}